\documentclass[12pt]{article}

\usepackage{amsmath}
\usepackage{amssymb}
\usepackage{dsfont} 
\usepackage{float}
\usepackage{mathtools}

\usepackage{graphicx}

\usepackage[dvipsnames]{xcolor}
\usepackage{color}

\usepackage{ulem} 

\newcommand{\be}{\begin{equation}}
\newcommand{\ee}{\end{equation}}
\newcommand{\bel}[1]{\begin{equation}\label{#1}}
\newcommand{\eeeee}{\end{equation}} 
\newcommand{\bea}{\begin{eqnarray}}
\newcommand{\eea}{\end{eqnarray}}
\newcommand{\balign}{\begin{align}}
\newcommand{\ealign}{\end{align}}
\newcommand{\ba}{\begin{array}}
\newcommand{\ea}{\end{array}}
\newcommand{\bfig}{\begin{figure}}
\newcommand{\efig}{\end{figure}}

\newcommand{\eref}[1]{(\ref{#1})}

\newcommand{\Fref}[1]{Fig.~\ref{#1}}

\allowdisplaybreaks

\newcommand{\exval}[1]{\mbox{$\langle \, {#1}\, \rangle$}}

\newcommand{\Prob}[1]{\mbox{${\rm Prob}\left[ \, {#1}\, \right]$}}

\newcommand{\rmd}{\mathrm{d}}
\newcommand{\rme}{\mathrm{e}}

\newcommand{\ddt}{\frac{\rmd}{\rmd t}}

\newcommand{\half}{\frac{1}{2}}

\newcommand{\one}{\mathds{1}}

\begin{document}

\title{A lattice gas model for generic one-dimensional Hamiltonian Systems}
\author{J. Schmidt$^{1, 2}$ \and G.M.~Sch\"utz$^{3}$ \and 
H. van Beijeren$^{4}$
}

\maketitle

{\small
\noindent $^{~1}$Bonacci GmbH, 
Robert-Koch-Str. 8, 50937 Cologne, Germany,
Email: schmidt@bonacci.de

\smallskip
\noindent $^{~2}$Institut f\"{u}r Theoretische Physik, 
Universit\"{a}t zu K\"{o}ln, Z\"ulpicher Str. 77, 50937 Cologne, Germany, 
Email: schmidt@thp.uni-koeln.de

\smallskip
\noindent $^{~3}$Institute of Biological Information Processing 5, Forschungszentrum J\"ulich, 52425 J\"ulich, Germany, 
Email: g.schuetz@fz-juelich.de 

\smallskip
\noindent $^{~4}$Institute for Theoretical Physics, 
Utrecht University, Leuvenlaan 4, 3584 CE, Utrecht, The Netherlands, 
Email: H.vanBeijeren@uu.nl\\[4mm]
}

\begin{abstract}
We present a three-lane exclusion process that exhibits the same universal
fluctuation pattern as generic one-dimensional Hamiltonian dynamics 
with short-range interactions, viz., with two sound modes
in the Kardar-Parisi-Zhang (KPZ) universality class (with dynamical 
exponent $z=3/2$ and symmetric Pr\"ahofer-Spohn scaling function) and a
superdiffusive heat mode with dynamical exponent $z=5/3$ and
symmetric L\'evy scaling function. The lattice gas model is
amenable to efficient numerical simulation. Our main findings,
obtained from dynamical Monte-Carlo simulation, are:
(i) The frequently observed numerical asymmetry of the sound modes 
is a finite time effect. (ii) The mode-coupling calculation of the scale
factor for the $5/3$-L\'evy-mode gives at least the right order of magnitude. 
(iii) There are significant diffusive corrections which are non-universal.
\end{abstract}

\newpage

\section{Introduction}

It is by now well-established that one-dimensional short-ranged 
Hamiltonian systems with conserved particle density, energy, momentum
generically exhibit universal time-dependent fluctuations of these quantities 
\cite{vanB12,Spoh14,Lepr16}. In the coordinate frame with zero 
center-of-mass velocity there are two oppositely moving sound modes.
These are in the Kardar-Parisi-Zhang (KPZ) universality class \cite{Halp15}
with dynamical exponent $z=3/2$ and, in a comoving frame, given by symmetric 
Pr\"ahofer-Spohn scaling function \cite{Prae04, Prae04_DATA}. Moreover, mode coupling 
theory predicts a zero-velocity heat mode with dynamical exponent $z=5/3$ 
with scaling form given by the $5/3$-stable symmetric L\'evy distribution 
\cite{vanB12,Spoh14}.

The fundamental assumptions underlying these predictions are that all slow variables 
of relevance for the long-time behavior of the time correlation functions 
are the long-wavelength Fourier components of the three conserved densities 
and that there are no further conservation laws. Thus one expects a broad universality
of dynamical non-equilibrium phenomena in one dimension.

It should be noted, however, that these assumptions are not sufficient to 
guarantee the KPZ/L\'evy/KPZ $(\frac{3}{2},\frac{5}{3},\frac{3}{2})$-scaling 
scenario. In fact, it has been shown that the KPZ and 5/3-L\'evy universality classes
are members of an infinite family of dynamical universality classes
whose dynamical exponents are -- quite remarkably -- given by the Kepler
ratios of neighbouring Fibonacci numbers, beginning with $z=2$ for
diffusion and including also the limiting value, which is the
golden mean $\varphi=(1+\sqrt{5})/2$ \cite{Popk15b,Popk16}\footnote{The first Kepler ratio with $z=1$ arises in conformally invariant dynamics of systems with long-range interactions \cite{Kare17}.}. Specifically, 
for three conservation laws one can have among these Fibonacci universality 
classes normal diffusion (Gaussian scaling function) or marginal
(logarithmic) superdiffusion ($z=2$), KPZ, modified KPZ 
(with unknown scaling function \cite{Spoh15}) and L\'evy 
(all with $z=3/2$), and L\'evy modes with $z=5/3$, $z=8/5$ and $z=\varphi$. 

Which combinations of these eight different Fibonacci universality classes 
arise depends on the structure of the mode coupling matrices that 
can be obtained from the macroscopic current-density relation
of the conserved densities \cite{Popk16, Spoh15}, and (only for marginal superdiffusion)
on the presence of a cubic term in the underlying fluctuating
hydrodynamic equation \cite{Devi92}. 
Moreover, notwithstanding the asymptotic predictions, one often
observes numerically significant deviations from the predicted scaling
behaviour which make the verification of the universality 
difficult in concrete applications \cite{Mend14,NaSch}
or even put non-equilibrium universality into question \cite{Hurt16}.

Here we address these issues by introducing a lattice gas model
for which mode coupling theory predicts the KPZ/L\'evy/KPZ-scaling
of generic Hamiltonian dynamics. One finds, without fine-tuning of parameters, 
the structure of the mode coupling matrices that is required
for the $(\frac{3}{2},\frac{5}{3},\frac{3}{2})$-scaling 
scenario, viz., non-vanishing self-coupling at quadratic order of the sound modes,
vanishing self-coupling of the heat mode, but non-vanishing coupling 
of the heat mode to the sound modes.

This model is a three-lane exclusion process which can be simulated
very efficiently, as detailed below. It turns out that deviations
from the scaling predictions are indeed strong for early times,
but there is a clear indication of convergence for larger times, as
discussed in the conclusions. The model allows for varying the
strength of the mode coupling coefficients and can thus be
employed to simulate scenarios that arise in other types of
models which are less amenable to numerical analysis.
In our model the conserved densities corresponding to the hydrodynamic 
densities of mass, momentum and energy are identified as linear combinations 
of the three particle densities in the three different lanes.
Ergodicity guarantees the absence of further conservation laws.

\section{The three-lane partial exclusion process}

\subsection{Definition}

Consider a three-lane asymmetric partial exclusion process with up to 
$m$ particles per site, $L$ sites per lane, and periodic boundary 
conditions. We denote by $n_k^{\lambda} \in \{0,1,\dots,m\}$ the occupation 
number on site $k$ on lane $\lambda$ with $\lambda \in\{-1,0,1\}$. Thus the 
time-dependent numbers $n_{k}^{\lambda}\left(t\right)$ represent the 
time evolution of a single realization of the stochastic process. 

The dynamics is Markovian. Particles hop randomly to nearest neighbour sites on 
the {\it same} lane with rates that depend not only on the occupation 
numbers of the departure site $k$ and the target site $k\pm 1$, but also on 
those of the neighbouring sites on the same and on the 
neighbouring lanes. The rates $r_k^{\lambda}$ for a particle jump from site 
$k$ to site $k+1$ and $\ell_k^{\lambda}$ from site $k$ to site $k-1$ are
\bea
\label{rminus}
r^{-}_k & = & \left\{ a_1 + \frac{b_1}{2} \left[n^{0}_k + n^{0}_{k+1}\right]  
+ \frac{d_1}{2} \left[n^{-}_{k-1} + n^{-}_{k+2}\right] \right\} n^{-}_k 
\left(1- \frac{n^{-}_{k+1}}{m}\right) \\
\label{lminus}
\ell^{-}_k & = & \left\{ a_2 + \frac{b_2}{2}\left[n^{0}_k + n^{0}_{k-1}\right] 
+ \frac{d_2}{2} \left[n^{-}_{k+1} + n^{-}_{k-2}\right] \right\} n^{-}_k 
\left(1 - \frac{n^{-}_{k-1}}{m}\right) \\
\label{r0}
r^{0}_k & = & 
\left\{ a_0 + \frac{b_1}{2} \left[n^{-}_k + n^{-}_{k+1}\right] 
+ \frac{b_2}{2} \left[n^{+}_k + n^{+}_{k+1}\right] \right\} 
n^{0}_k \left(1 - \frac{n^{0}_{k+1}}{m}\right) \\
\label{l0}
\ell^{0}_k & = & 
\left\{ a_0 + \frac{b_1}{2} \left[n^{+}_k + n^{+}_{k-1}\right] 
+ \frac{b_2}{2} \left[n^{-}_k + n^{-}_{k-1}\right] \right\} 
n^{0}_k \left(1 - \frac{n^{0}_{k-1}}{m}\right) \\
\label{rplus}
r^{+}_k & = & \left\{ a_2 + \frac{b_2}{2} \left[n^{0}_k + n^{0}_{k+1}\right] 
+ \frac{d_2}{2} \left[n^{+}_{k-1} + n^{+}_{k+2}\right] \right\} 
n^{+}_k \left(1- \frac{n^{+}_{k+1}}{m}\right) \\
\label{lplus}
\ell^{+}_k & = & \left\{ a_1 + \frac{b_1}{2}\left[n^{0}_k + n^{0}_{k-1}\right] 
+ \frac{d_1}{2} \left[n^{+}_{k+1} + n^{+}_{k-2}\right] \right\} 
n^{+}_k \left(1 - \frac{n^{+}_{k-1}}{m}\right).
\eea
The parameter range is $a_i \geq 0$, $b_i+d_i \geq - a_i/m$, 
$b_1+b_2 \geq - a_0/m$ to ensure positivity of all jump rates. 
The inter-lane coupling strength is given by the 
constants $b_{1,2}$ and we shall assume at least one of them to be non-zero to have
interaction between the lanes.
Complete or partial decoupling also takes place if one of the lanes is empty or completely filled. 
We exclude these trivial cases from our considerations. The case $m=1$ 
corresponds to coupled exclusion processes.
Due to the periodic boundary conditions the total number of particles 
$N_\lambda = \sum_k n_{k}^{\lambda}$ in each lane is conserved.

\subsection{Steady state properties}

\subsubsection{Stationary distribution}

For parameters $z_{\lambda} \geq 0$ the product measures with site marginals
\bel{marggen}
\Prob{n_k^{\lambda} = n} = 
\frac{z_{\lambda}^n}{(1+z_{\lambda})^m} {m \choose n}
\ee
are a family of stationary distributions which one proves 
by observing that both the right-hopping process and the 
left-hopping process individually leave \eref{marggen} 
invariant, which comes from a cancellation of the terms
$$\half \left[b_2 \left(n^{0}_k + n^{0}_{k+1}\right)
+ d_2 \left(n^{+}_{k-1} + n^{+}_{k+2}\right) \right] 
\left(n^{+}_k - n^{+}_{k+1}\right)$$
and
$$ \half b_2 \left(n^{+}_k + n^{+}_{k+1}\right)
\left(n^{0}_k - n^{0}_{k+1}\right)$$
in the lattice sum over $k$, and similarly for the other hopping terms. 
Notice that the exclusion parameter $m$ does not appear in these equations. 

The fugacity $z_{\lambda}$ parametrizes the density 
\bel{density}
\rho_{\lambda} = m \frac{z_{\lambda}}{1+z_{\lambda}} 
\ee
on lane $\lambda$ obtained from \eref{marggen}.
The total particle number $N_{\lambda}$ in each lane is conserved under 
the dynamics, but is a fluctuating quantity among realizations in the grand canonical
ensemble defined by the invariant measure \eref{marggen}. The product form 
yields the diagonal density covariance matrix $K$ with matrix elements
\bel{Kij}
K_{\lambda\mu} = z_\mu 
\frac{\partial \rho_\lambda}{\partial z_\mu} = \kappa_{\lambda} \delta_{\lambda\mu}
\ee
with
\bel{kappa}
\kappa_{\lambda}=\rho_{\lambda}/(1+z_{\lambda})=\rho_{\lambda}(1-\rho_{\lambda}/m).
\ee
Expectations of time-independent functions $f$ in the stationary distribution are denoted by
$\exval{f}$.

\subsubsection{Currents}

The so-called {\it instantaneous} currents for the bond $(k,k+1)$
in lane $\lambda$ are the functions 
\bel{instcurr}
j^{\lambda}_k(t) := r^{\lambda}_{k}(t) - \ell^{\lambda}_{k+1}(t)
\ee
of the time-dependent occupation numbers $n^{\lambda}_k(t)$ which represent a single
realization of the stochastic process. 
For an arbitrary initial distribution $P_0$ of the particles, these instantaneous currents yield the microscopic continuity equations for the time-dependent expected local density $\rho^{\lambda}_k(t):= \exval{n^{\lambda}_k(t)}_{P_0}$ as
\bel{continuity}
\ddt \rho^{\lambda}_k(t) = \exval{j^{\lambda}_{k-1}(t)}_{P_0} - 
\exval{j^{\lambda}_k(t)}_{P_0}.
\ee
Here the brackets denote averages over histories of the
process and an arbitrary initial distribution $P_0$ of the particles.
For the stationary currents $j_\lambda$, which are translation invariant, 
one has $j_{\lambda} = \exval{j^{\lambda}_k}$.
The product structure of the stationary distribution yields
\bea
\label{jplusgen}
j_{-} & = & 
- \left( a + b \rho_0 + d \rho_{-} \right) \rho_{-} 
\left(1-\frac{1}{m} \rho_{-}\right) 
\\
\label{j0gen}
j_0 & = & 
b\left(\rho_{+}-\rho_{-}\right)  \rho_0 \left(1-\frac{1}{m} \rho_{0}\right) 
\\
\label{jminusgen}
j_{+} & = &  
\left( a + b \rho_0 + d \rho_{+} \right) \rho_{+} 
\left(1-\frac{1}{m} \rho_{+}\right) 
\eea
with
\bel{def:abd}
a := a_2 - a_1, \quad b := b_2 - b_1, \quad d := d_2 - d_1.
\ee
The subscript here corresponds to the superscript in the previous equations for the
instantaneous current.
Notice that the stationary currents depend only on the difference of the
individual rates $a_{1,2},b_{1,2},d_{1,2}$ and are independent of $a_0$.
The exclusion parameter $m$ only renormalizes the densities, the asymmetry
parameter $a$ and the overall time scale.

From the stationary current density relation \eref{jplusgen} - 
\eref{jminusgen} one obtains the current Jacobian $J$ with matrix elements 
\bel{Jacobiandef}
J_{\lambda\mu} = \frac{\partial j_\lambda}{\partial \rho_\mu}.
\ee

We are specifically interested in the symmetric case $\rho_+=\rho_-=:\rho$.
In this case 
the current Jacobian takes the form
\bel{Jacobisym}
J =  \left( \ba{ccc}
- \delta & - b\kappa & 0 \\
- b\kappa_0 & 0 &  b\kappa_0 \\
0 &  b\kappa & \delta
\ea \right)
\ee
with
\bea
\label{def:delta}
& & \delta = a + b \rho_0 + 2\left(d - \frac{a+b\rho_0}{m}\right) \rho -
\frac{3d}{m} \rho^2, \\
\label{def:kappa}
& & \kappa = \rho \left(1- \frac{\rho}{m}\right), \quad \kappa_0 = \rho_0 \left(1- \frac{\rho_0}{m}\right).
\eea

The Hessians $H^\nu$ are defined by the matrix elements
\bel{Hessdef}
H^\nu_{\lambda\mu} = 
\frac{\partial^2 j^\nu}{\partial \rho_\lambda \partial \rho_\mu}.
\ee
Defining
\bel{Hessentries}
x = d-\frac{a+b\rho_0 + 3d \rho}{m}, \quad y = b\left(1 - \frac{2\rho}{m}\right),
\quad z = b\left(1 - \frac{2\rho_0}{m}\right)
\ee
one has for equal densities $\rho_\pm=\rho$
\bel{Hesssym}
H^{-} = - \left( \ba{ccc} 
2x & y & 0 \\
y & 0 & 0 \\
0 & 0 & 0
\ea \right), \,\,
H^0 = z \left( \ba{ccc} 
0 & -1 & 0 \\
-1 & 0 & 1 \\
0 & 1 & 0
\ea \right), \,\,
H^{+} = \left( \ba{ccc} 
0 & 0 & 0 \\
0 & 0 & y \\
0 & y & 2x
\ea \right) .
\ee
By construction, the Hessians are symmetric. 

\subsubsection{Dynamical structure function}

The (real-space) dynamical structure function is the matrix $\bar{S}_k(t)$ 
with matrix elements
\bel{dynstruc}
\bar{S}^{\lambda\mu}_k(t) = \exval{(n^{\lambda}_k(t)-\rho_\lambda) 
(n^{\mu}_0(0)-\rho_\mu)}.
\ee
Here the brackets without subscript denote an average over histories of the process
and the stationary distribution.
Because of the factorized stationary distribution one has
\bel{dynstrucstat}
\bar{S}^{\lambda\mu}_k(0) = \kappa_\lambda \delta_{\lambda,\mu} \delta_{k,0}
\ee
and, due to particle number conservation,
\begin{equation}
\sum_{k=0}^{L-1} \bar{S}_k\left(t\right) = K  \quad \forall t \geq 0
\label{eq:compressibility_identity}
\end{equation}
with the compressibility matrix defined in \eref{Kij}.
From the continuity equation \eref{continuity}, the absence of stationary correlations,
and the ensemble property \eref{Kij} one also has for the infinite
lattice the exact property
\begin{equation}
\ddt \sum_k k \bar{S}_k\left(t\right) = JK \quad \forall t \geq 0
\end{equation}
where $J$ is the current Jacobian \eref{Jacobiandef}. The matrix product $JK$ is symmetric as has been 
proved in \cite{Gris11}, thus linking the purely static
compressibility $K$ with the 
dynamics encoded in $J$. 

The dynamical structure function describes the flow and broadening
of density fluctuations in the steady state. To elucidate this property
we diagonalize the current Jacobian $J$, i.e., we study 
$RJR^{-1}= V =: \text{diag}\left(v_-,v_0,v_+\right)$ with eigenvalues $v_\alpha$. 
The three eigenvectors of $J$ define the real-space eigenmodes 
\bel{eigenmode_position} 
\phi_k^\alpha := \sum_\mu R_{\alpha\mu} (n_k^\mu - \rho_\mu)
\ee
which are fluctuation fields that travel with velocities $v_\alpha$.
We normalize the transformation $R$ by
\bel{Rnorm}
R K R^{T} = \one 
\ee
where $K$ is the compressibility matrix defined in \eref{Kij}. 

With the transformation matrix $R$ to normal modes, 
one obtains the dynamical structure function for the eigenmodes
\bel{Seigen} 
S^{\alpha\beta}_{k}\left(t\right)
= \left(R\bar{S}_{k}\left(t\right)R^{T}\right)_{\alpha\beta}
= \exval{\phi^\alpha_k(t) \phi^\beta_0}.
\ee
Thus, with the normalization \eref{Rnorm} the transformed dynamical structure function
$S$ satisfies 
\begin{equation}
\sum_{k} S_k\left(t\right) = \one, \quad 
\ddt \sum_k k S_k\left(t\right) = V.
\end{equation}
The first relation normalizes the ``mass'' of each mode
and the second relation gives the propagation velocities
of the modes. We stress that these are exact relations valid on the infinite lattice for all finite times $t\geq 0$.

The lattice Fourier transform $\hat{\bar{S}}(p,t)$ of the
dynamical structure function is defined by
\bel{dynstrucFT}
\hat{\bar{S}}^{\lambda\mu}(p,t) = \sum_{k=0}^{L-1} \rme^{-2\pi i pk} \bar{S}^{\lambda\mu}_k(t),
\ee
with $p$ of the form $p=m/L$ with $m$ integer.
We can also consider the correlation between the Fourier-transformed density fields
\be 
\hat{n}^{\lambda}(p,t) = \sum_{k=0}^{L-1} \rme^{-2\pi i pk} (n^{\lambda}_k(t)-\rho_\lambda),
\ee
from which one constructs
\be 
\tilde{\bar{S}}^{\lambda\mu}(p,q,t) = \exval{\hat{n}^{\lambda}(p,t)\hat{n}^{\mu}(q,0)}.
\ee
Exploiting translation invariance one finds $\tilde{\bar{S}}(p,q,t) 
= L \hat{\bar{S}}(p,t) \delta_{p+q,0}$ and therefore
\bel{structure_fourier_p_q_t} 
\hat{\bar{S}}(p,t) = \frac{1}{L} \tilde{\bar{S}}(p,-p,t).
\ee
Fourier eigenmodes are calculated similar to \eref{eigenmode_position} as
\bel{eigenmode_fourier} 
\hat{\phi}^\alpha(p,t) := \sum_\mu R_{\alpha\mu} \hat{n}^\mu(p,t)
\ee
and from \eref{structure_fourier_p_q_t} one concludes
\bel{Seigen_fourier} 
\hat{S}^{\alpha\beta}\left(p, t\right)
= \left(R\hat{\bar{S}}\left(p, t\right)R^{T}\right)_{\alpha\beta}
= \frac{1}{L}\exval{\hat{\phi}^\alpha(p,t) \hat{\phi}^\beta(-p,0)}
\ee
satisfying
\be
\hat{S}(0,t) = \one.
\ee

\subsection{Monte Carlo simulations for the structure function}

\subsubsection{Monte Carlo algorithm}

Monte Carlo simulations are performed for a periodic system of 
very large length $L=10^6$ with fixed particle numbers $N_\lambda$. 
The densities are then given by $\rho_\lambda = N_\lambda/L$.
Since the exclusion parameter $m$ is immaterial from a theoretical
perspective and we look for optimal numerical efficiency we choose full exclusion corresponding to $m=1$. 

Notice that on switching to a system with fixed particle numbers, the 
stationary distribution modifies and each configuration with $N_\lambda$
occupied sites on lane $\lambda$ becomes equally likely to be observed. After 
the drawing of an initial configuration from this uniform stationary distribution 
the system is evolved in time by using random sequential update.

Making use of translation invariance and ergodicity, we define the
Monte-Carlo estimator for the structure functions
\begin{eqnarray}
\bar{\sigma}_{L,k}^{\lambda\mu}\left(M,\tau,t\right) & = & 
\frac{1}{M}\sum_{j=1}^{M}\frac{1}{L}\sum_{l=1}^{L}\left[
n_{l}^{\lambda}\left(j\tau\right)n_{l+k}^{\mu}
\left(j\tau+t\right)-\rho_{\lambda}\rho_{\mu}\right]
\label{eq:structure-function-ESTIMATOR}
\end{eqnarray}
where  $l+k$ has to be taken modulo $L$. Further, 
$\tau$ is the time between measurements and $M$ is
the total number of measurements.
In order to compute the structure functions we generate
$P$ independent initial configurations
yielding Monte Carlo estimators $\bar{\sigma}^{(p)}_{L,k}$ 
for each initial configuration. Averaging over the
initial configurations then yields the numerical structure function
\bel{eq:structure-function-ESTIMATOR-av}
S^{\text{MC}}_{L,k}\left(P,M,\tau,t\right) = \frac{1}{P}\sum_{p=1}^{P}
R\bar{\sigma}_{L,k}^{\left(p\right)}\left(M,\tau,t\right)R^{T}
\ee
which depends on the simulation parameters $L,P,M,\tau$, on the
space and time parameters $k,t$, and on the model parameters.
On choosing $P$ and $M$ sufficiently large the Monte Carlo error for 
the numerical structure function becomes Gaussian distributed 
with zero mean and  standard deviation scaling as $\sim 1/\sqrt{PM}$. 
Our values
for these simulation parameters are given in Sec.~\eref{Sec:Results}.

\subsubsection{Canonical ensemble correction for finite-size effects}
\label{Subsec:FScorrections}

In the grand canonical ensemble the dynamical structure function vanishes rapidly 
at finite time $t$ outside the "light cone" which is the region enclosed by the 
modes with the lowest and highest velocity. For numerical purposes, we define 
the light cone by the 
interval
\bel{lightcone}
 \mathbb{L}= [v_lt-c_lt^{1/z_l}, v_ht+c_ht^{1/z_h}],
\ee
 where $v_l~(v_h)$ is the lowest (highest) mode velocity, $z_l~(z_h)$  is the dynamical exponent of the  corresponding mode and
$c_l~(c_h)$ is a constant chosen such that correlations outside are indeed
vanishing within statistical measurement precision, after taking care of the following remark.

For a finite system the constant numbers of particles $N_{\lambda}=L
\rho_{\lambda}$ introduce long-range correlations extending over the
whole lattice even for $t=0$. The canonical
invariant measure is uniform, which for $m=1$ yields the static structure function
\bel{ssfcan} 
\exval{n^\lambda_k n^\mu_0}_{N_+,N_0,N_-} - \rho_{\lambda}\rho_{\mu} = 
- \frac{1}{L-1} \rho_{\lambda} (1-\rho_{\lambda}) \delta_{\lambda\mu} 
\quad \mbox{ for } k \neq 0
\ee
for fixed particle numbers $N_{\lambda}$.
Thus in a canonical ensemble at time $t>0$ and $k$ outside the light cone one has, within numerical measurement precision,
\be 
\bar{S}_{\text{can},k}^{\lambda\mu}\left(t\right) = - \frac{\rho_{\lambda}\left(1-\rho_{\lambda}\right)}{L-1} \delta_{\lambda\mu} .
\ee
Transforming to eigenmodes one gets 
\be 
S_{\text{can},k}^{\alpha\beta}\left(t\right) = -\frac{1}{L-1} \delta_{\alpha\beta}.
\ee
Thus the structure function in the canonical ensemble has a constant offset of order $L^{-1}$ 
compared to the grandcanonical structure function used above to describe correlations in the infinite system 
and which does not require this correction. 

For the numerical precision of our Monte-Carlo data
this offset is relevant and needs to be taken into account.
Correcting for the finite size effects exposed in \eref{ssfcan} we will use the quantities
\be 
\tilde{\sigma}_{L,k}^{\lambda\mu}\left(M,\tau,t\right) = \bar{\sigma}_{L,k}^{\lambda\mu}\left(M,\tau,t\right) 
+ \frac{\rho_{\lambda}\left(1-\rho_{\lambda}\right)}{L-1} \delta_{\lambda\mu}.
\ee

However, other finite size effects do appear in either ensemble. First of all, at times $t>L/(v_h-v_l)$ 
different modes will meet and add interactions over which we have no control. At even longer times, 
proportional to $L^{z_\lambda}$, the discreteness of the allowed set of values for the Fourier 
variable $p$ becomes noticeable in the various mode coupling contributions. To avoid having 
to deal with this we will restrict our comparisons between simulations and theory to times 
shorter than this. The system size is chosen as large as numerically possible, 
but large enough to allow for neglecting within Monte-Carlo accuracy, the finite size effects
discussed above.

\subsubsection{Choice of rates}

We recall that only the rate differences $a,b,d$ defined in \eref{def:abd}
occur in the currents and therefore in the mode-coupling matrices. 
For efficient simulation, avoiding irrelevant jump processes, 
we choose the jump rates of the three-lane model as follows:
\begin{eqnarray}
a_{0} & = & -\min\left(0,b\right)\\
a_{1} & = & -\min\left(0,b\right)-\min\left(0,d\right)\\
a_{2} & = & a+a_{1}\\
b_{1} & = & 0\\
b_{2} & = & b\\
d_{1} & = & 0\\
d_{2} & = & d
\end{eqnarray}
When choosing the rates in this way, the hopping rates simplify to
\begin{eqnarray}
l_{k}^{-} & = & \left\{ a_{2}+\frac{b}{2}\left(n_{k}^{0}+n_{k-1}^{0}\right)+\frac{d}{2}\left(n_{k+1}^{-}+n_{k-2}^{-}\right)\right\} n_{k}^{-}\left(1-n_{k-1}^{-}\right)\label{eq:hopping_rates_start}\\
r_{k}^{-} & = & a_{1}n_{k}^{-}\left(1-n_{k+1}^{-}\right)\\
l_{k}^{0} & = & \left\{ a_{0}+\frac{b}{2}\left(n_{k}^{-}+n_{k-1}^{-}\right)\right\} n_{k}^{0}\left(1-n_{k-1}^{0}\right)\\
r_{k}^{0} & = & \left\{ a_{0}+\frac{b}{2}\left(n_{k}^{+}+n_{k+1}^{+}\right)\right\} n_{k}^{0}\left(1-n_{k+1}^{0}\right)\\
l_{k}^{+} & = & a_{1}n_{k}^{+}\left(1-n_{k-1}^{+}\right)\\
r_{k}^{+} & = & \left\{ a_{2}+\frac{b}{2}\left(n_{k}^{0}+n_{k+1}^{0}\right)+\frac{d}{2}\left(n_{k-1}^{+}+n_{k+2}^{+}\right)\right\} n_{k}^{+}\left(1-n_{k+1}^{+}\right)\label{eq:hopping_rates_end}
\end{eqnarray}
and result in less storage accesses during the simulation. Additionally, for $a\geq 0$
this choice guarantees to have all rates greater than or equal to 0 and we avoid having
the processes $r_{k}^{-}$ and $l_{k}^{+}$ for $a_{1}=0$.

Notice that we have chosen the rates such that $l^+=r^-$; $ l^-=r^+$;  $\rho_+=\rho_-$ 
and the rates $r^0$ and $l^0$ depend on the occupations of the neighboring lanes in a symmetric way.
This way we are guaranteed to find the velocities in the form $v_{\pm}=\pm v$ and $v_0=0$, 
just like for the sound modes and the heat mode in one dimensional Hamiltonian systems. 
Below we will use these terms to refer to the modes in our model system as well.

\section{Predictions from mode coupling theory}

In generic Hamiltonian dynamics as well as in the present model with our choices of parameters 
all three velocities $v_\alpha$ are 
different \cite{vanB12} and this will be used throughout the discussion in this section. 
Under these conditions one expects the off-diagonal elements of the 
dynamical structure function \eref{Seigen} to decay fast
and on large scales one is left with the diagonal terms $S_\alpha(x,t)$ with continuous space coordinate $x$.

We follow our previous work \cite{Popk16} and use the mode coupling equations
\be
\label{ModecouplingRS}
\partial_t S_{\alpha}(x,t) =  -v_\alpha \partial_x + D_{\alpha} \partial_x^2
 +  \int_0^t \rmd s \int_{-\infty}^\infty \rmd y\,
 S_{\alpha}(x-y,t-s) M_{\alpha\alpha}(y,s)
\ee
with memory term
\bel{MemoryRS}
M_{\alpha\alpha}(y,s) =
2 \sum_{\beta,\gamma} (G^{\alpha}_{\beta\gamma})^2
\partial_y^2 S_\beta(y,s)S_\gamma(y,s) 
\ee
to predict the large scale-scale behaviour of $S_\alpha(x,t)$.
Here the $G^{\alpha}$ are the mode coupling matrices obtained from the
Hessians $H^\alpha$ \eref{Hessdef} 
by the transformation
\bel{G}
G^{\alpha} = 
\half \sum_\beta R_{\alpha\beta} \left(R^{-1}\right)^T H^\beta R^{-1}.
\ee
The diffusion coefficients $D_\alpha$ turn out to be immaterial for the theoretical
predictions that arise from \eref{ModecouplingRS}.
By construction, the mode coupling matrices are symmetric and related to the
mode coupling matrices $W^{\alpha}$ of \cite{vanB12} by
$W^{\alpha} = 2 G^{\alpha}$.

\subsection{Dynamical scaling}

In the scaling limit $t\to\infty$ and $x\to\infty$ with constant scaling 
variable $u_\alpha = E_\alpha (x-v_\alpha t)t^{-1/z_\alpha}$ with a 
non-universal scale parameter $E_\alpha$, non-universal mode velocity 
$v_\alpha$ and universal dynamical exponent $z_\alpha$ the 
mode coupling equations \eref{ModecouplingRS} with memory term 
\eref{MemoryRS} can be solved exactly for all modes $\alpha$ 
and any number of conservation laws \cite{Popk16}. Thus one can determine 
from them self-consistently for each mode the dynamical exponent 
$z_\alpha$ and the corresponding scaling form of the dynamical structure 
function $S_\alpha(x,t) = (E_\alpha t)^{-1/z_\alpha} s_\alpha(u)$. 

We note that generally, in the case of distinct mode velocities ($v_\alpha \not = v_\beta$ 
for all $\alpha\not = \beta$) and  the absence of purely
diffusive modes $\delta$ (for which $G^\delta_{\alpha\alpha} = 0$
for all modes $\alpha$), any mode $\alpha$ with non-vanishing 
self-coupling coefficient $G^\alpha_{\alpha\alpha}$ is expected to be
in the KPZ universality class with dynamical exponent $z_\alpha = 3/2$.
Rather than determining the corresponding scaling functions by the 
mode coupling equations \eref{ModecouplingRS} we use in our approach 
the exact Pr\"ahofer-Spohn scaling $s_\alpha(u) = f_{PS} (u)$ \cite{Prae04}. 

For Hamiltonian dynamics this argument applies to
the two sound modes with velocities $v_\pm = \pm v$ 
so that their asymptotic behavior is expected to become
\bel{eq:KPZ-Mode-Spohn-Asym-Solution}
S_{\pm}\left(x,t\right) \simeq  \left(\lambda t\right)^{-2/3}\cdot
f_{\text{PS}}\left(\left(x\mp vt\right)\cdot\left(\lambda t\right)^{-2/3}\right)
\ee
with 
\bel{def:lambda}
\lambda = 2^{3/2}\left|G_{++}^{+}\right| = 2^{3/2}\left|G_{--}^{-}\right|
\ee
and speed $v = |v_\pm| \neq 0$.
The scaling function $f_{\text{PS}}$ is known exactly and satisfies
\begin{eqnarray}
f_{\text{PS}}\left(-x\right) & = & f_{\text{PS}}\left(x\right)\\
\max f_\text{PS}(x)&=&f(0)=0.542461\ldots\\
\intop_{-\infty}^{\infty}f_{\text{PS}}\left(x\right)\text{d}x & = & 1\\
\intop_{-\infty}^{\infty}\left(f_{\text{PS}}\left(x\right)\right)^{2}
\text{d}x & = & 0.389813\ldots=:c_{\text{PS}} .
\end{eqnarray}
There is no expression in closed form for $f_\mathrm{PS}$ which, however, has been calculated 
with high precision and is tabulated in \cite{Prae04_DATA}. Using this data a more precise 
calculation of $c_\mathrm{PS}$ can be found in Eq.~(74) of \cite{Popk16}.

In a setup with three modes, two symmetric KPZ-modes traveling with $v_+=-v_-=v$, 
identical self-coupling $|G^+_{++}|=|G^-_{--}|$, and a mode 0 with vanishing self-coupling 
$G^0_{00}=0$ and mode-velocity, but non-vanishing symmetric coupling 
$|G^0_{++}|=|G^0_{--}|$ to the KPZ modes, Eq. \eref{ModecouplingRS} predicts dynamical exponent $z_0 = 5/3$
for this mode. The corresponding scaling function is then a symmetric $5/3$-stable L\'evy distribution
given by
\bel{Heat_Scalinng_Fct_Position_Space}
S_{0}(x,t) = \frac{1}{2\pi}\intop_{-\infty}^{\infty}\exp\left(-E_{0}t\left|p\right|^{5/3}\right)\rme^{ipx}\text{d}p
\ee
with \cite{Popk16}
\bel{eq:heat_scaling_constant}
E_{0} =\frac{a_h}{2} v^{-1/3} \left|G_{++}^{0}\right|^{2} 
\left|G_{++}^{+}\right|^{-2/3}.
\ee
and 
\be 
a_{h} = 4\Gamma\left(\frac{1}{3}\right)\sin\left(\frac{\pi}{3}\right)c_{\text{PS}} = 3.6175\ldots
\ee
Here $\Gamma(\cdot)$ is the Gamma function.
At this point we would like to emphasize that the mode coupling solution for the heat mode differs from \cite{vanB12}  only in the prefactor $a_{h,\cite{vanB12}}=1.6712$.

We recall that non-linear fluctuating hydrodynamics \cite{Spoh14}
predicts the presence of diffusive corrections to these asymptotic
results.
Defining the Fourier transform as
\bel{FT}
\hat{f}\left(p\right) = 
\intop_{-\infty}^{\infty}f\left(x\right)\rme^{-ipx}\text{d}x
\ee
the Fourier representation of the asymptotic heat scaling function
with diffusive correction included is given by 
\bel{eq:Heat-Scaling_Fct_DIFF_FINITE_TIME}
\intop_{-\infty}^{\infty}S_{0}\left(x,t\right)
\rme^{-ipx}\text{d}p \simeq  \exp\left(-E_{0}t\left|p^{5/3}\right| - D_{0}p^{2}t\right).
\ee
Here $D_0$ is a phenomenological constant.

\subsection{Eigenmodes and mode coupling matrices of the three-lane model}

The scenario described above is expected for generic Hamiltonian dynamics
with short-range interactions. Specifically, one requires in the
frame of vanishing center-of-mass velocity
\be 
v_0 = 0, \quad v_\pm = \pm v
\ee
and the mode coupling symmetry \cite{vanB12}
\bel{Gsym}
G^\gamma_{\alpha\alpha} = - G^{-\gamma}_{-\alpha-\alpha}.
\ee
with $G^-_{--} \neq 0$ and $G^0_{--}\neq 0$.
Hence, to verify these requirements, one needs to compute all diagonal
mode coupling coefficients $G^\alpha_{\beta\beta}$ for the three-lane
exclusion process.

Indeed, the characteristic polynomial of the Jacobian \eref{Jacobisym} yields the eigenvalues
\bel{eigenvalsym}
v_0 = 0, \quad v_{\pm} = \pm v
\ee
with the strictly positive constant
\bel{vsym}
v = \sqrt{\delta^2 + 2 b^2 \kappa \kappa_0} .
\ee

We choose the diagonalizing matrix $R$ such that
\be 
R J R^{-1} = \mbox{diag} (-v,0,v).
\ee
Together with the normalization \eref{Rnorm} this yields
\bel{Rsym2}
R = \frac{1}{\xi} \left( \ba{ccc} 
\frac{b\sqrt{\kappa_0}}{v-\delta} & \frac{1}{\sqrt{\kappa_0}} & -\frac{b\sqrt{\kappa_0}}{v+\delta} \\
\frac{1}{\sqrt{\kappa}} & -\frac{\delta}{b\sqrt{\kappa}\kappa_0} & \frac{1}{\sqrt{\kappa}} \\
-\frac{b\sqrt{\kappa_0}}{v+\delta} & \frac{1}{\sqrt{\kappa_0}} &  \frac{b\sqrt{\kappa_0}}{v-\delta}
\ea \right), \,
R^{-1} = \frac{1}{\xi} \left( \ba{ccc} 
 \frac{b\sqrt{\kappa_0}\kappa}{v-\delta} & \sqrt{\kappa} & -\frac{b\sqrt{\kappa_0}\kappa}{v+\delta} \\
\sqrt{\kappa_0} & -\frac{\delta}{b\sqrt{\kappa}} & \sqrt{\kappa_0} \\
-\frac{b\sqrt{\kappa_0}\kappa}{v+\delta} & \sqrt{\kappa} &  \frac{b\sqrt{\kappa_0}\kappa}{v-\delta}
\ea \right)
\ee
where
\bel{def:xi}
\xi = \frac{v}{b\sqrt{\kappa\kappa_0}} = \sqrt{2+\frac{\delta^2}{b^2\kappa\kappa_0}} > 0.
\ee
Notice the following symmetry
\bel{Rsymmetry}
R_{\alpha\beta} = R_{-\alpha-\beta}, \quad (R^{-1})_{\alpha\beta} = (R^{-1})_{-\alpha-\beta}
\ee
which will play a role below.

With the sum
\bea
\label{Hess2}
\tilde{H}^{\gamma} & = & \sum_{\lambda} R_{\gamma \lambda} H^{\lambda} \\
& = & 
\left( \ba{ccc}
- 2xR_{\gamma -} & -(zR_{\gamma 0}+yR_{\gamma -}) & 0 \\
-(zR_{\gamma 0}+yR_{\gamma -}) & 0 & zR_{\gamma 0}+yR_{\gamma +} \\
0 & zR_{\gamma 0}+yR_{\gamma +} & 2xR_{\gamma +}
\ea \right).
\eea
of the Hessians \eref{Hesssym} we find
\bea
G^\gamma_{\alpha\alpha} & = & \frac{1}{2} 
\sum_{\mu\nu} (R^{-1})_{\mu\alpha} \tilde{H}^\gamma_{\mu\nu}
(R^{-1})_{\nu\alpha} \\
& = & \frac{1}{2} \left[ \tilde{H}^\gamma_{--} \left((R^{-1})_{-\alpha}\right)^2 
+ \tilde{H}^\gamma_{++} \left((R^{-1})_{+\alpha}\right)^2 \right] \nonumber \\
& & +  \tilde{H}^\gamma_{0-} (R^{-1})_{0\alpha} (R^{-1})_{-\alpha}
+ \tilde{H}^\gamma_{+0} (R^{-1})_{+\alpha} (R^{-1})_{0\alpha} \\
& = & x \left[ R_{\gamma+} \left((R^{-1})_{+\alpha}\right)^2 -
R_{\gamma-} \left((R^{-1})_{-\alpha}\right)^2 \right] \nonumber \\
& & + (R^{-1})_{0\alpha} \left[ (z R_{\gamma0} + yR_{\gamma+}) (R^{-1})_{+\alpha} 
- (z R_{\gamma0} + y R_{\gamma-}) (R^{-1})_{-\alpha} \right] .
\eea
The mode coupling symmetry \eref{Gsym} is indeed satisfied without any
fine-tuning of parameters, as can be seen as follows. 

The Hessians \eref{Hesssym} exhibit the symmetry
\bel{Hsymmetry}
H^\gamma_{\alpha\beta} = - H^{-\gamma}_{-\alpha-\beta}.
\ee
which induces $\tilde{H}^\gamma_{\alpha\beta} = - \tilde{H}^{-\gamma}_{-\alpha-\beta}$.
Therefore, with $(R^{-1})^T_{\alpha\beta}=R^{-1}_{\beta\alpha}$ one gets
\bea
G^\gamma_{\alpha\alpha} 
& = & \frac{1}{2} 
\sum_{\mu\nu} (R^{-1})_{\mu\alpha} \tilde{H}^\gamma_{\mu\nu} (R^{-1})_{\nu\alpha} \\
& = & - \frac{1}{2} 
\sum_{\mu\nu} (R^{-1})_{\mu\alpha} \tilde{H}^{-\gamma}_{-\mu,-\nu} (R^{-1})_{\nu\alpha} \\
& = & - \frac{1}{2} 
\sum_{\mu\nu} (R^{-1})_{-\mu,\alpha} \tilde{H}^{-\gamma}_{\mu\nu} (R^{-1})_{-\nu,\alpha} \\
& = & - \frac{1}{2} 
\sum_{\mu\nu} (R^{-1})_{\mu,-\alpha} \tilde{H}^{-\gamma}_{\mu\nu} (R^{-1})_{\nu,-\alpha} \\
& = & - G^{-\gamma}_{-\alpha,-\alpha} 
\eea

One finds for the independent non-vanishing coefficients
\bea
\label{---}
G^-_{--} & = & - \frac{\kappa\kappa_0}{v^2} \sqrt{\kappa}
\left[ x \left(\frac{\delta^2}{\kappa\kappa_0} + \frac{b^2}{2}\right) 
+ \frac{y b \delta }{\kappa} + z b^2 \right] \\
G^-_{00} & = & - \frac{\kappa\kappa_0}{v^2} b \sqrt{\kappa}
\left( x b - \frac{y\delta}{\kappa}\right) \\
G^-_{++} & = & - \frac{\kappa\kappa_0}{v^2} 
b^2 \sqrt{\kappa}
\left( \frac{x}{2} - z \right) \\
G^0_{--} & = & - \frac{\kappa\kappa_0}{v^2} b \sqrt{\kappa_0}
\left((x-z)\frac{\delta}{\kappa_0} + y b \right).
\eea
Some steps of the lengthy computation are presented in the appendix.

\section{Monte Carlo results for the structure-function}
\label{Sec:Results}

\subsection{Data fit using $L_{1}$-distance}

The prediction for the heat mode 
and the exact Pr\"ahofer-Spohn scaling function involve the
the non-universal scale parameters $\lambda$ (given in \eref{def:lambda}) and 
$E_0$ (given in \eref{eq:heat_scaling_constant}). In order
to allow for a comparison with Monte-Carlo results we define 
analogously to \cite{Mend14} the fitted scaling parameters
$\lambda^{\text{fit}}_{\alpha}$
as the minimum of the $L_{1}$-distance
\begin{equation}
\lambda_{\alpha}^{\text{fit}} = \underset{\lambda}{\text{argmin}}
\left[\sum_{k\in\mathcal{K}}\left|S_{\alpha}\left(k,t\right)
-\left(\lambda t\right)^{-1/z_{\alpha}}f_{\alpha}^{\text{th}}
\left(\left(\lambda t\right)^{-1/z_{\alpha}}\left(k-v_{\alpha}t\right)\right)
\right|\right]
\end{equation}
where $\mathcal{K}$ is the set of calculated sampling points.

For a model independent comparison of simulation results and theory we consider 
analogously to \cite{Mend14} the scale factor coefficients
\begin{align}
a_{s}& =\frac{\lambda_{+}}{\left|G_{++}^{+}\right|}=2^{3/2}=2.82842\ldots\\
a_{s}^{\text{fit}} & =\frac{\lambda_{\text{+}}^{\text{fit}}}{\left|G_{++}^{+}\right|}\\
a_{h} & =\frac{2E_{0}}{v^{-1/3}\cdot\left|G_{++}^{0}\right|^{2}\cdot\left|G_{++}^{+}\right|^{-2/3}}
=4\Gamma\left(\frac{1}{3}\right)\sin\left(\frac{\pi}{3}\right)c_{\text{PS}}
=3.61751\ldots\\
a_{h}^{\text{fit}} & =\frac{2E_{\text{0,fit}}}{v^{-1/3}\cdot\left|G_{++}^{0}\right|^{2}\cdot\left|G_{++}^{+}\right|^{-2/3}}.
\end{align}

\subsection{Overview}

With Monte Carlo simulation we aim at exploring the asymptotic behavior
in numerically accessible times. Therefore, to guarantee a quickly
vanishing mode overlap, the model parameters $a,b,d$ and the
particle densities $\rho,\rho_0$ should be chosen such that
the speed $v$ of the KPZ sound modes becomes as
large as possible. Furthermore, the scale factors 
$\lambda$, $E_0$ for the sound and heat modes should be as large as possible so as to
dominate any diffusive ($z=2$) finite time contribution. 
However, 
according to (\ref{eq:heat_scaling_constant}) a strong
KPZ selfcoupling $G^+_{++}$ and large sound velocity $v$
weakens the $5/3$-L\'evy-mode
scale factor $E_0$. Thus, to access the asymptotic regime of both modes
we have to find a system with well balanced scale factors for both
modes.

Below we present two examples showing the following properties:
\begin{enumerate}
\item Good KPZ-modes and a $5/3$-L\'evy-mode with diffusive
finite time effects.
\item Good $5/3$-L\'evy-mode and KPZ-modes with diffusive finite
time effects.
\end{enumerate}
In both examples the convergence to the predicted asymptotic results
becomes apparent after subtraction of the diffusive correction.

\subsection{Simulation data}

\subsubsection{Example 1: Good KPZ-modes and a $5/3$-L\'evy-mode with diffusive finite time effects.}

We choose the model parameters
\begin{eqnarray}
\rho & = & 0.25\\
\rho_{0} & = & 0.3\\
a & = & 0.5031056\\
b & = & 0.4968944\\
d & = & 0
\end{eqnarray}
From this we obtain the sound velocities $v_{\pm} = \pm 0.35465$
and the mode coupling matrices
\begin{eqnarray}
G^{-} & = & \left(\begin{matrix}0.2139 & 0.1269 & -0.0255\\
0.1269 & 0.0509 & -0.0154\\
-0.0255 & -0.0154 & 0.0176
\end{matrix}\right)\\
G^{0} & = & \left(\begin{matrix}0.0854 & 0 & 0\\
0 & 0 & 0\\
0 & 0 & -0.0854
\end{matrix}\right)\\
G^{+} & = & \left(\begin{matrix}-0.0176 & 0.0154 & 0.0255\\
0.0154 & -0.0509 & -0.1269\\
0.0255 & -0.1269 & -0.2139
\end{matrix}\right).
\end{eqnarray}
The theoretical scale parameters are thus given by
\be 
\lambda = 0.605, \quad E_{0} = 5.209\cdot10^{-2}.
\ee

\begin{figure}[H]
\begin{centering}
\includegraphics[width=10cm]{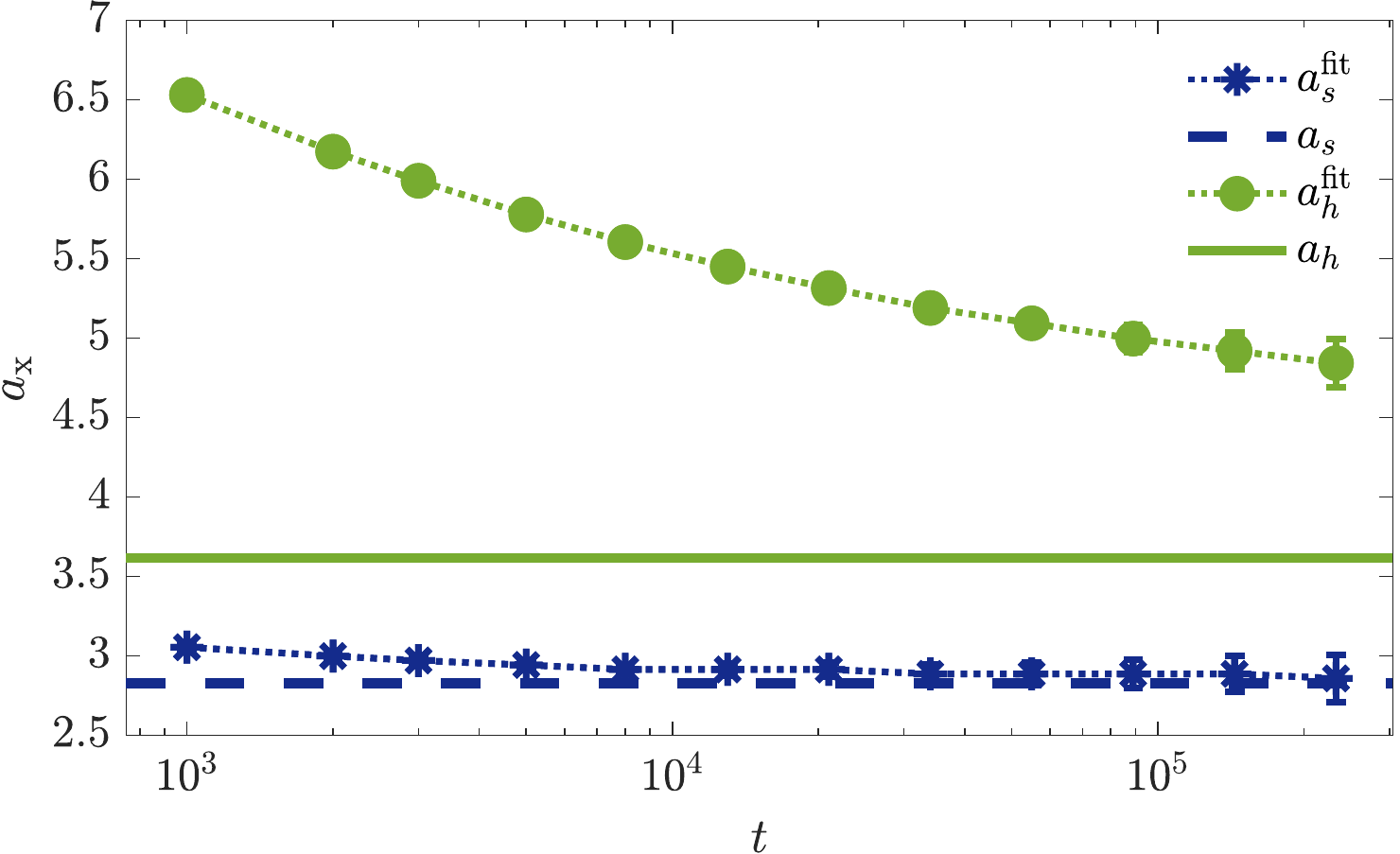}
\par\end{centering}
\caption{\label{Fig1}
Comparison of the numerical scale factors $a_{\pm,0}$ with their theoretical values.
Convergence of the measured scale parameter $a^{\mathrm{fit}}_+$
and theoretical asymptotic value. Asterisks and broken horizontal
line: KPZ sound modes,
Bullets and full horizontal line: Heat mode. The dotted lines between
data points are guides to the eye.}
\end{figure}

As Monte Carlo parameters defined in \eref{eq:structure-function-ESTIMATOR}
and \eref{eq:structure-function-ESTIMATOR-av} we choose
\begin{eqnarray}
P & = & 1000\\
M & = & 200\\
\tau & = & 1000.
\end{eqnarray}
We first consider the fit of the scale parameters obtained from the
$L_1$ distance, as shown in Fig.~\ref{Fig1}. 
We observe a monotone convergence of the numerical $L_1$ distance and 
of the scale parameters $\lambda_{\pm,0}$ and $a_{s,h}$
resp. to their theoretical values. At the largest time 
$t=2.33 \cdot 10^5$ the scale parameter $a_{s}^{\text{fit}}$
of the KPZ sound mode is very close to the exact theoretical value 
(precision $\approx 1\%$).
The scale parameter $a_{h}^{\text{fit}}$ of the heat mode, however,
is significantly off ($\approx 34\%$) the theoretical prediction.

In order to explore the heat mode further we make a fit in Fourier space
and include a diffusive correction. In this way we find
\begin{eqnarray}
E_{0,\mathrm{fit}} & = & \left(1.07\pm 0.02\right) E_{0}\\
D_{0,\mathrm{fit}} & = & 0.95\pm 0.006
\end{eqnarray}
where $D_{0,\mathrm{fit}}$ is the fitted diffusion coefficient of the
heat mode. This result
corresponds to a deviation of only $\approx 7\%$ from the
theoretical result. This indicates that diffusive
corrections are significant. However, since $t$ is not
sufficiently large to be in the asymptotic regime one cannot
tell from this result whether the $7\%$ difference between 
numerics and theory is due to residual finite-time corrections
or to an imprecision of the mode-coupling approximation.

\begin{figure}[H]
\centering{}\includegraphics[width=10cm]{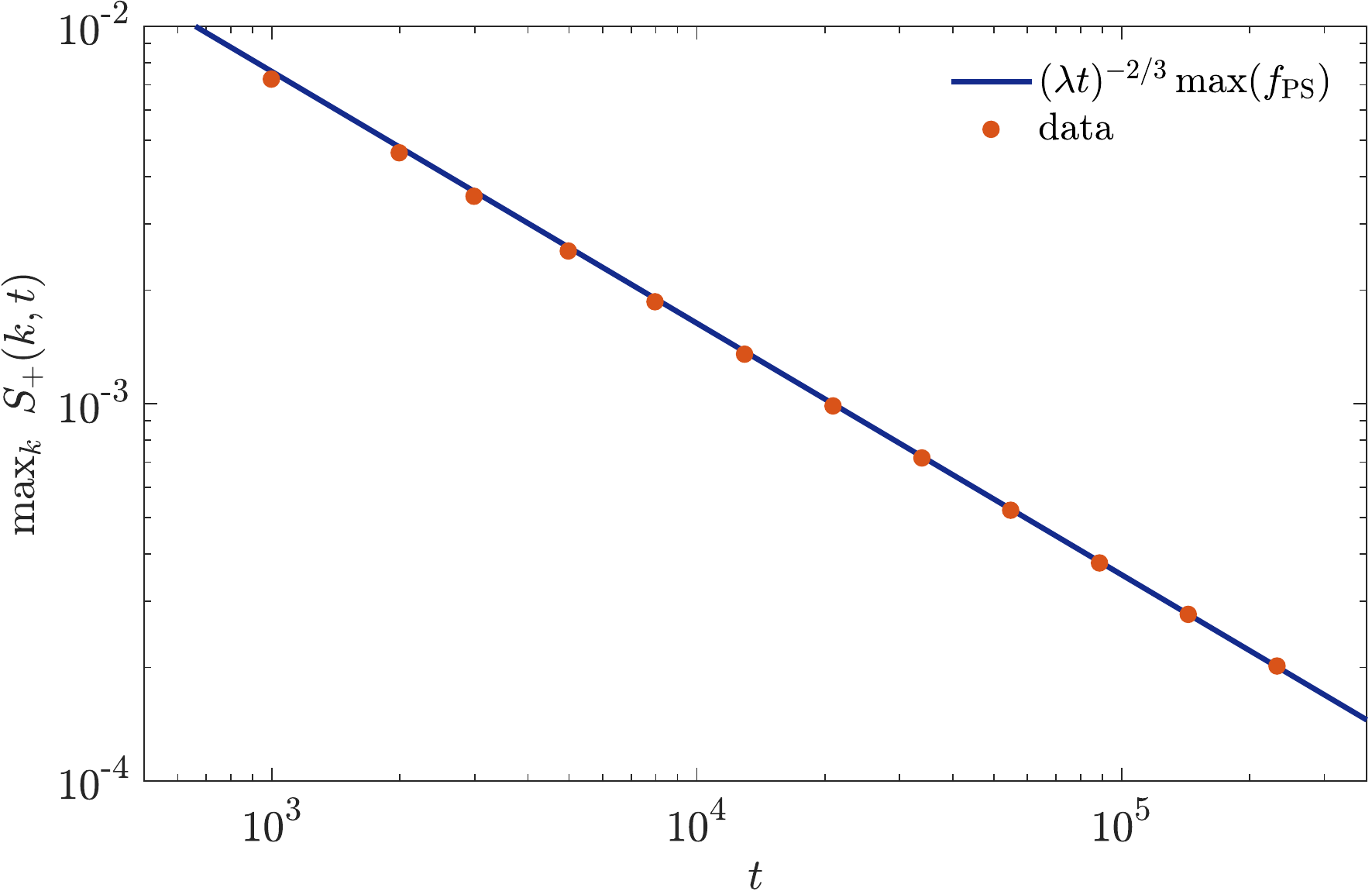}
\caption{\label{fig:EX1_log-log-plot-forKPZ-MODE}
Log-log plot for the 
value of the KPZ-mode
maximum (data points with error bars) 
versus time compared to the asymptotic solution (\ref{eq:KPZ-Mode-Spohn-Asym-Solution}) (full curve). Statistical errors are in order of symbol size.}
\end{figure}

\begin{figure}[H]
\centering{}\includegraphics[width=10cm]{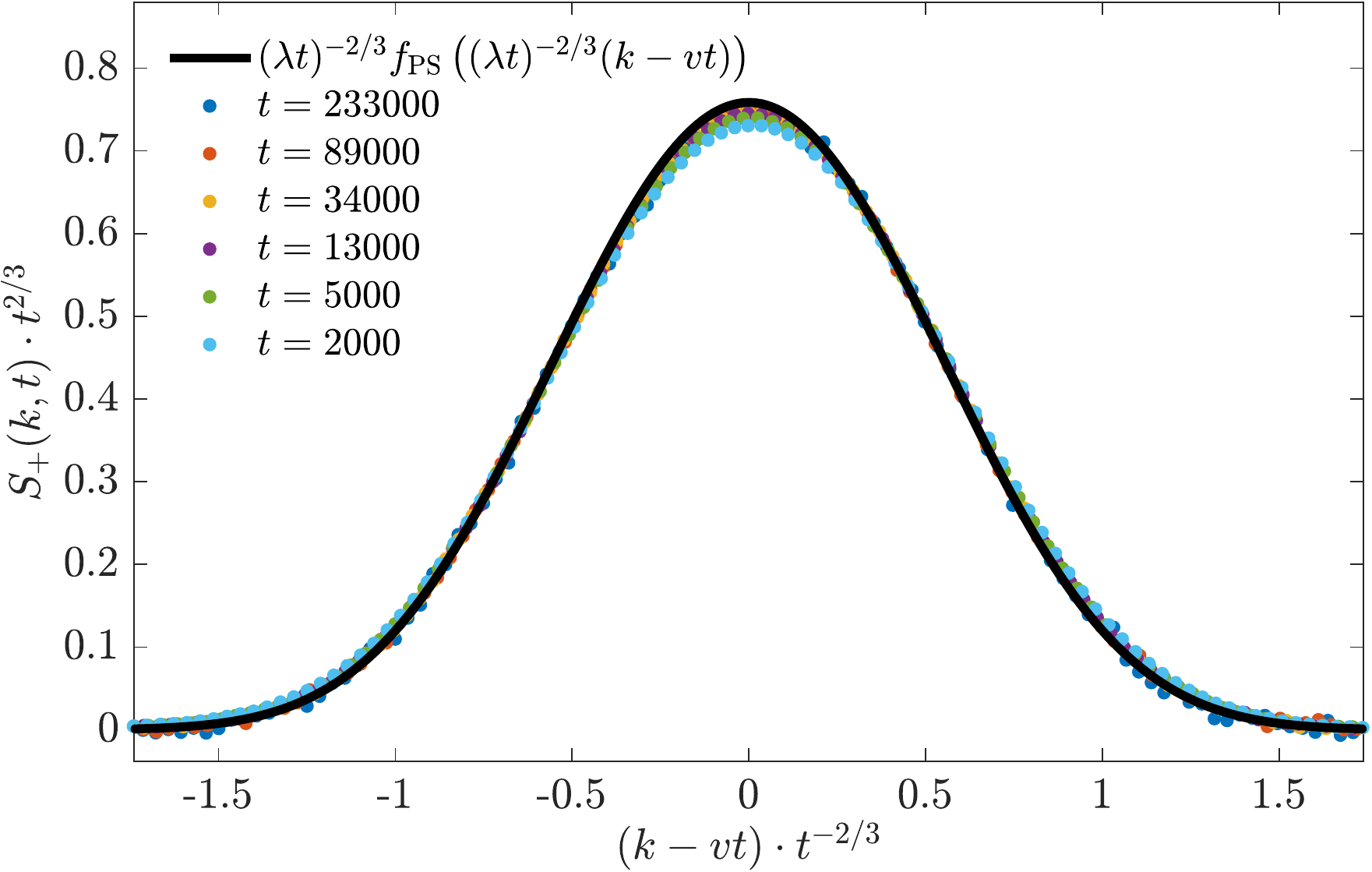}
\caption{\label{fig:EX1_Data-collapse-for-KPZ-mode}
Data collapse in position
space of the KPZ-mode compared for various times 
to the asymptotic Pr\"ahofer-Spohn
scaling-function (\ref{eq:KPZ-Mode-Spohn-Asym-Solution}) (Black curve). Statistical errors are of the order of symbol size.}
\end{figure}

For a more detailed analysis we plot the value of the maximum of the
sound modes together with the exact theoretical value 
(\Fref{fig:EX1_log-log-plot-forKPZ-MODE})
as well as a scaling plot of the full dynamical structure function in 
position space (\Fref{fig:EX1_Data-collapse-for-KPZ-mode}). One finds excellent convergence of the numerical
data to the theoretical curve.

\begin{figure}[H]
\begin{centering}
\includegraphics[height=4.75cm]{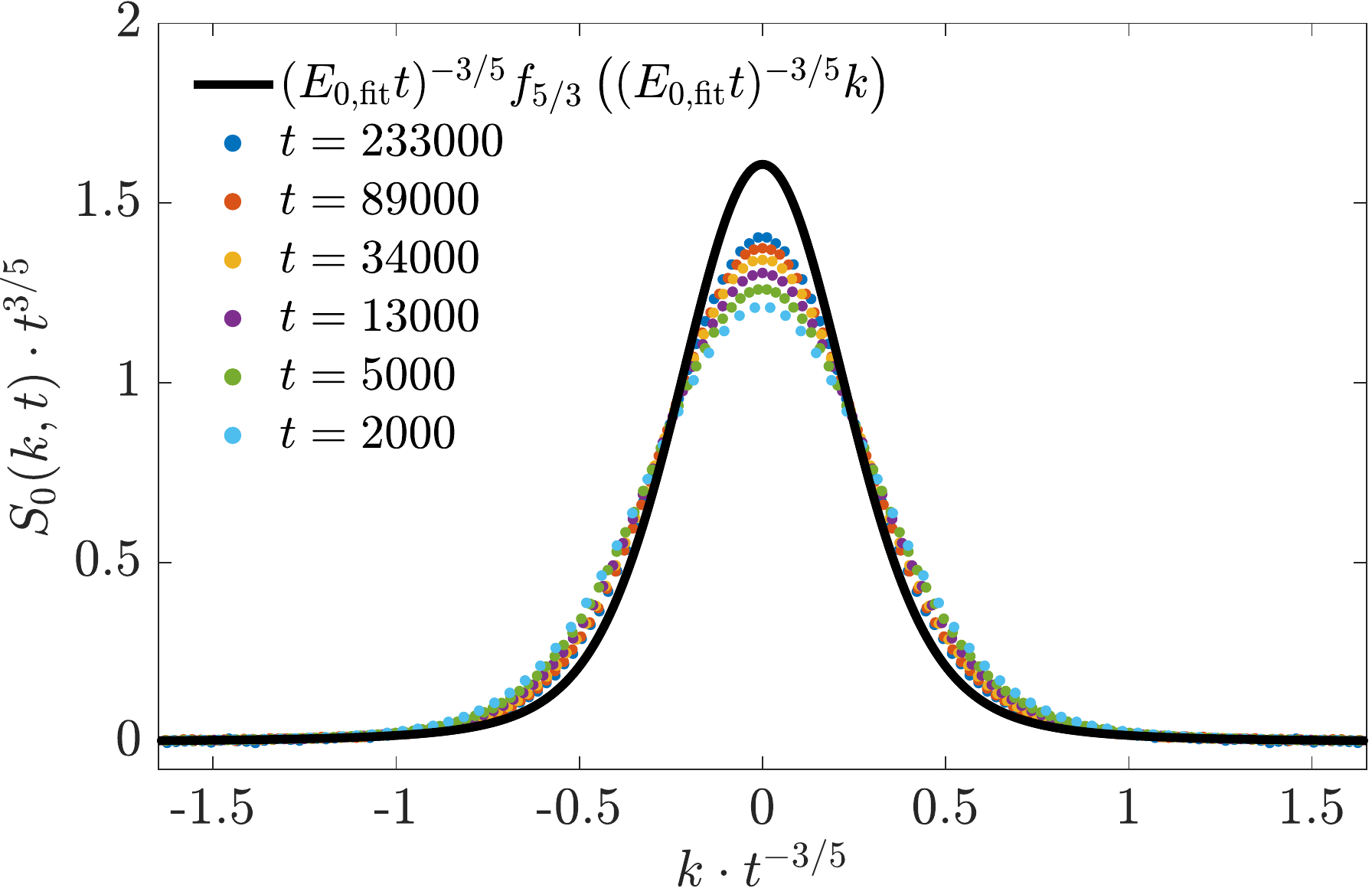}
\includegraphics[height=4.75cm]{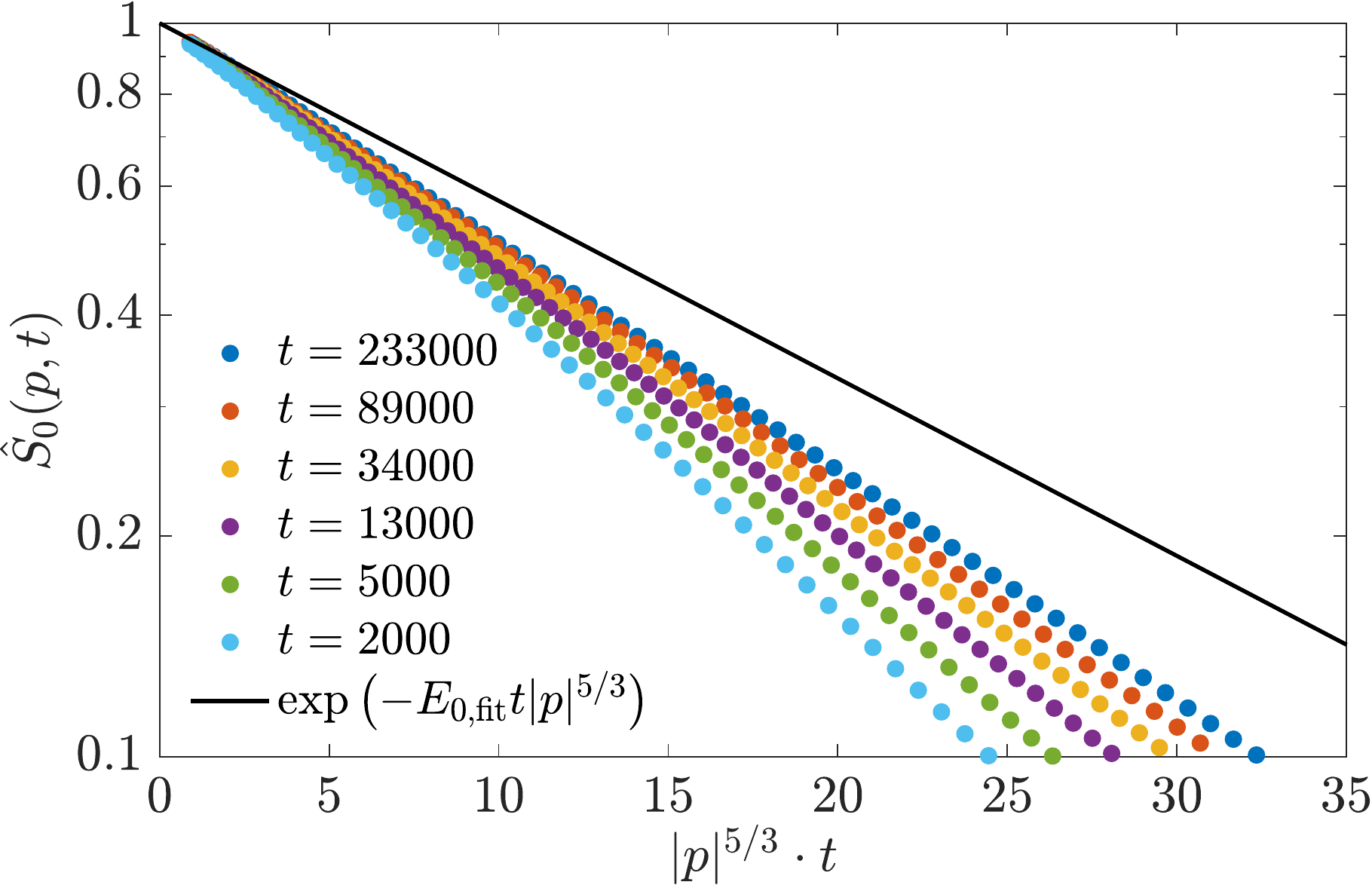}
\par\end{centering}
\caption{\label{fig:EX1_Data-collapse-for-HEAT-mode}
Data collapse of the $5/3$-L\'evy-mode and scaling-exponent $z=5/3$ compared
to a fitted symmetric $5/3$-L\'evy-stable distribution described
in Fourier representation by eq. (\ref{eq:Heat-Scaling_Fct_DIFF_FINITE_TIME})
and (\ref{eq:heat_scaling_constant}) with $E_{0,\mathrm{fit}}=\left(1.07\mp0.02\right)\cdot E_{0}$
and $E_{0}= 5.209\cdot10^{-2}$. Left panel: Position
space, Right panel: Momentum space. 
Statistical errors are in order of symbol
size.}
\end{figure}

For the heat mode the measured dynamical structure function does not
collapse in the time range accessible to simulation, see 
\Fref{fig:EX1_Data-collapse-for-HEAT-mode}, both for position space
and momentum space.

\begin{figure}[H]
\begin{centering}
\includegraphics[width=12cm]{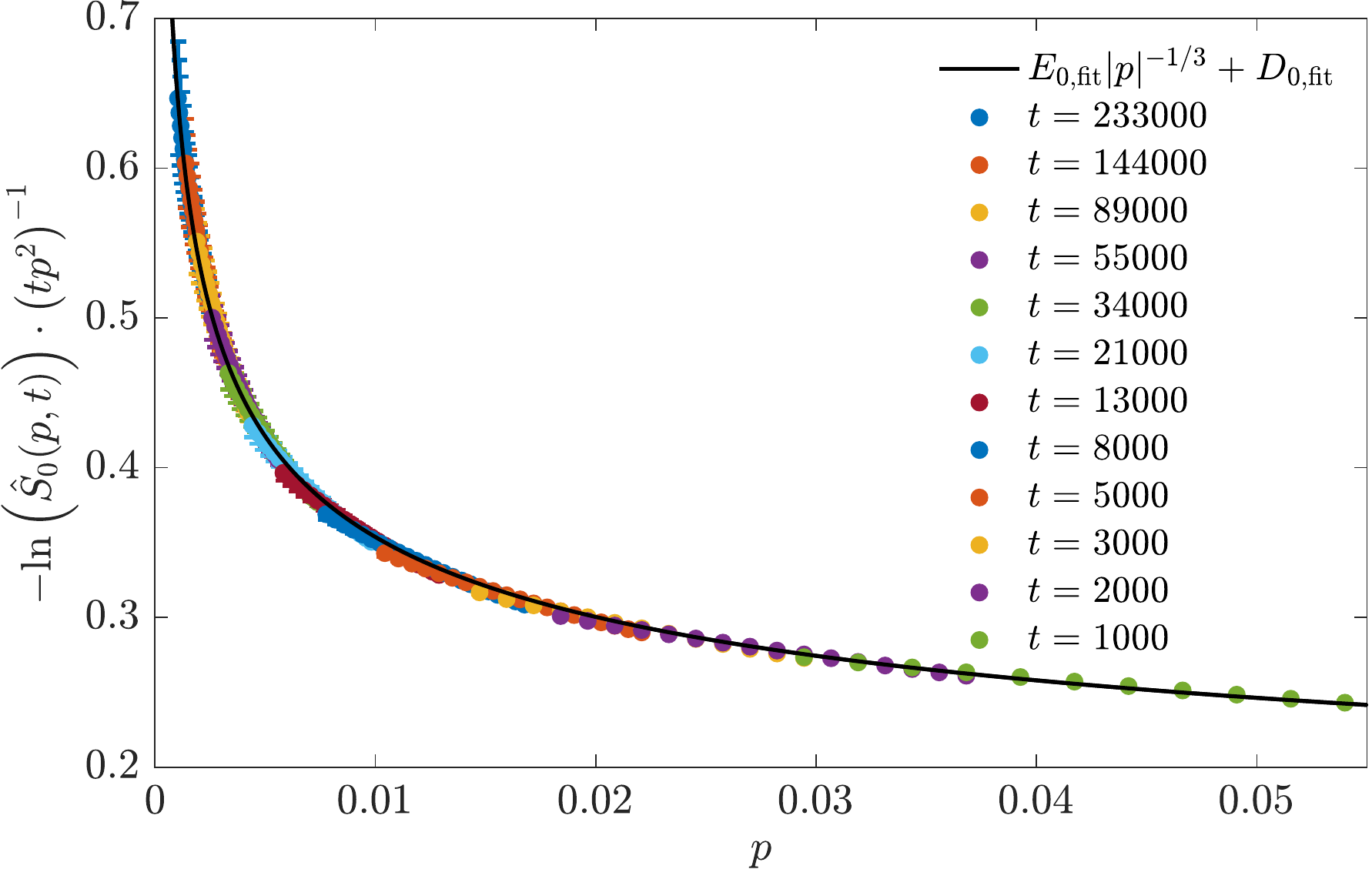}
\par\end{centering}
\caption{\label{fig:EX1_Heat_Mode_diffusive_finite_time_fit}
Fit for diffusive
finite time effects in the $5/3$-L\'evy-mode in fourier-space. Data
is compared to a fitted symmetric $5/3$-stable distribution including
diffusive finite time effects given by eq. (\ref{eq:Heat-Scaling_Fct_DIFF_FINITE_TIME})
with $D_{0,\mathrm{fit}}=0.095\pm0.006$, $E_{0,\mathrm{fit}}=\left(1.07\mp0.02\right)\cdot E_{0}$
and $E_{0}= 5.209\cdot10^{-2}$.}
\end{figure}

The absence of scaling suggests the presence of strong diffusive corrections.
This is confirmed by inclusion of a diffusive correction in the Fourier
transform, as shown in \Fref{fig:EX1_Heat_Mode_diffusive_finite_time_fit}.
The data collapse is excellent.

\subsubsection{Example 2: Good $5/3$-L\'evy-mode and KPZ-modes with diffusive finite time effects}

Here we demonstrate that the presence of diffusive corrections is
{\it not} a particular feature of the heat mode. They appear also in the
sound modes and their impact has to do with the
values of the scale parameters and hence with
the mode coupling coefficients. To this end we take as model parameters
\begin{eqnarray}
\rho & = & 0.395\\
\rho_{0} & = & 0.09\\
a & = & 0.522066265\\
b & = & 0.477933735\\
d & = & 0.
\end{eqnarray}

From this we obtain $v_{\pm} = \pm 0.1517$ and
\begin{eqnarray}
G^{-} & = & \left(\begin{matrix}0.1487 & 0.1342 & -0.0318\\
0.1342 & 0.0635 & -0.0358\\
-0.0318 & -0.0358 & 0.064
\end{matrix}\right)\\
G^{0} & = & \left(\begin{matrix}0.1556 & 0 & 0\\
0 & 0 & 0\\
0 & 0 & -0.1556
\end{matrix}\right)\\
G^{+} & = & \left(\begin{matrix}-0.064 & 0.0358 & 0.0318\\
0.0358 & -0.0635 & -0.1342\\
0.0318 & -0.1342 & -0.1487
\end{matrix}\right)
\end{eqnarray}
and
\be 
\lambda = 0.4205, \quad
E_{0} = 0.2927.
\ee
Notice, in comparison to the first example, the smaller value
of the sound mode self coupling coefficients $|G^\pm_{\pm\pm}|$ and
the much larger value $|G^0_\pm|$ of the heat mode coupling coefficient
which lead to correspondingly different values of 
$\lambda$ and $E_{0}$.

\begin{figure}[H]
\begin{centering}
\includegraphics[width=10cm]{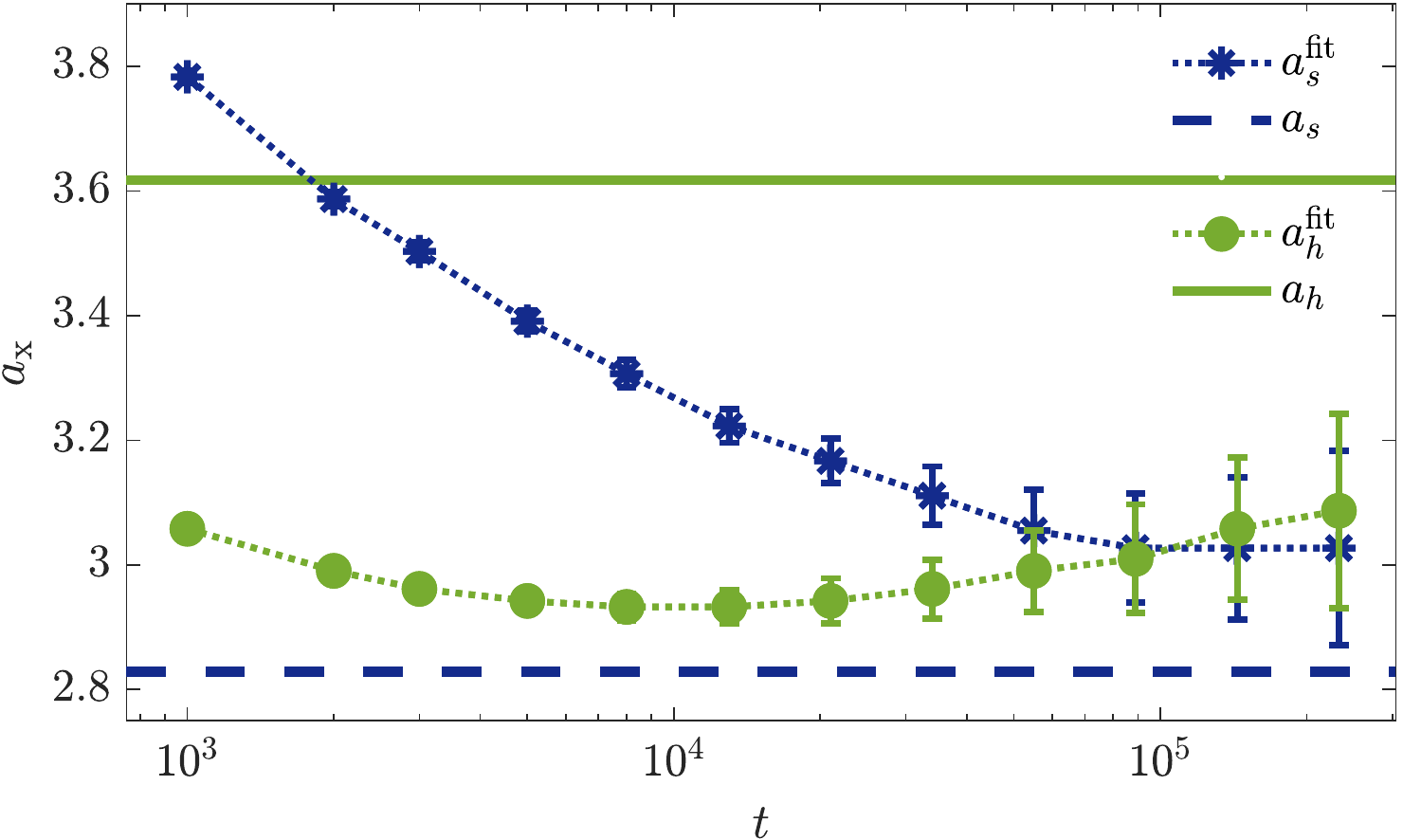}
\par\end{centering}
\caption{\label{zeta2}
	Comparison of the numerical scale factors $a_{\pm,0}$ with their theoretical values.
	Convergence of the measured scale parameter $a^{\mathrm{fit}}_+$
	and theoretical asymptotic value. Asterisks and broken horizontal
	line: KPZ sound modes,
	Bullets and full horizontal line: Heat mode. The dotted lines between
	data points are guides to the eye.
}
\end{figure}

As Monte Carlo parameters we choose
\begin{eqnarray}
P & = & 1000\\
M & = & 200\\
\tau & = & 1000.
\end{eqnarray}
One finds as in example 1 a convergence of the scale
parameters to the theoretical
asymptotic values (\Fref{zeta2}). However, at the
largest time $t=233000$ there are still significant deviations from 
the asymptotic values, both for the KPZ sound mode ($\approx 6\%$) and the
L\'evy heat mode ($\approx 14\%$ from below). The correction to
$a_0$ for the heat mode is non-monotonic.

\begin{figure}[H]
\centering{}\includegraphics[width=10cm]{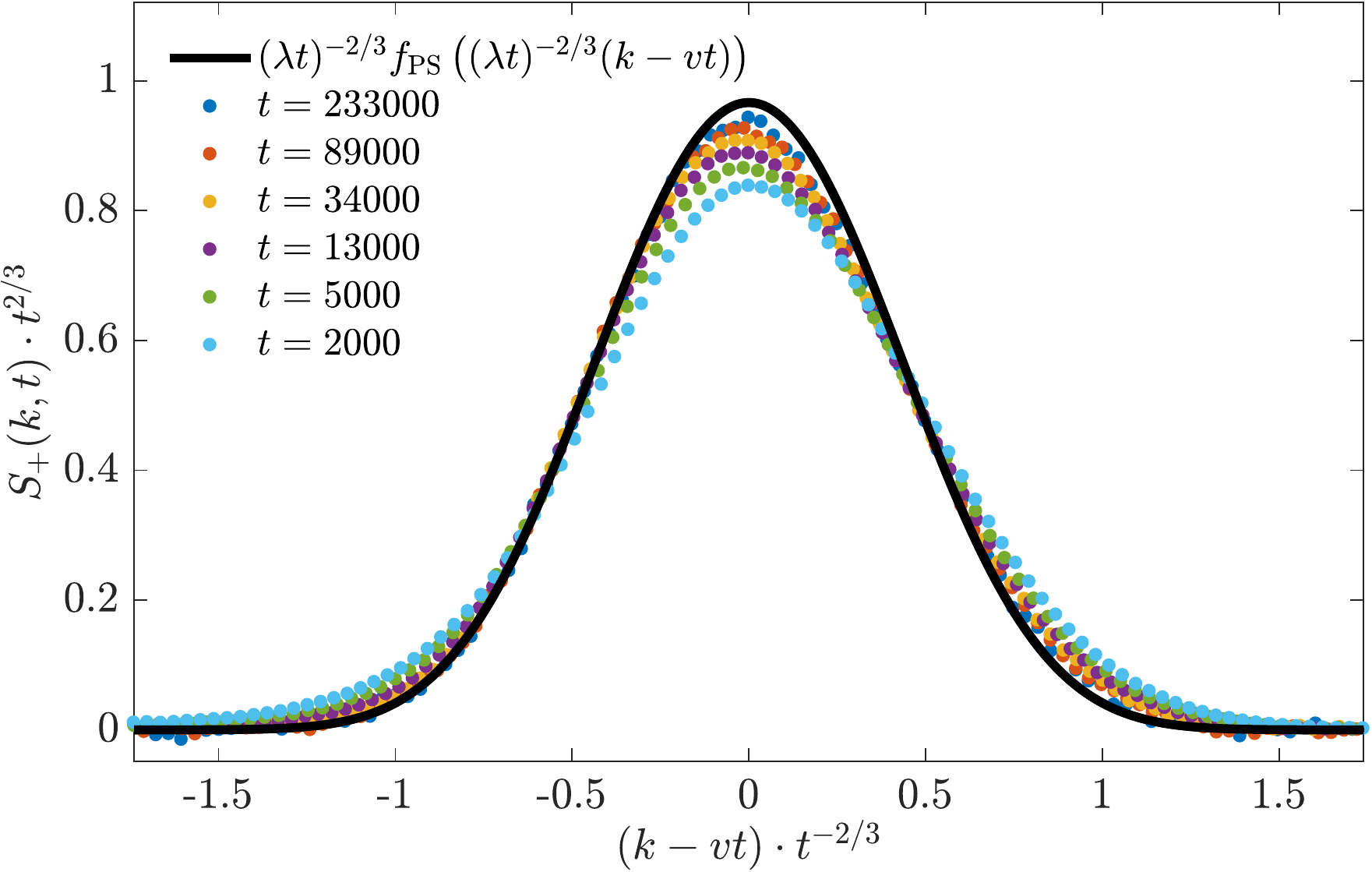}
\caption{\label{fig:EX2_Data-collapse-for-KPZ-mode} 
Data collapse in position
space of the KPZ sound mode compared to the asymptotic Pr\"ahofer-Spohn
scaling-function. Statistical errors are in order of symbol size.}
\end{figure}

\begin{figure}[H]
\centering{}\includegraphics[width=10cm]{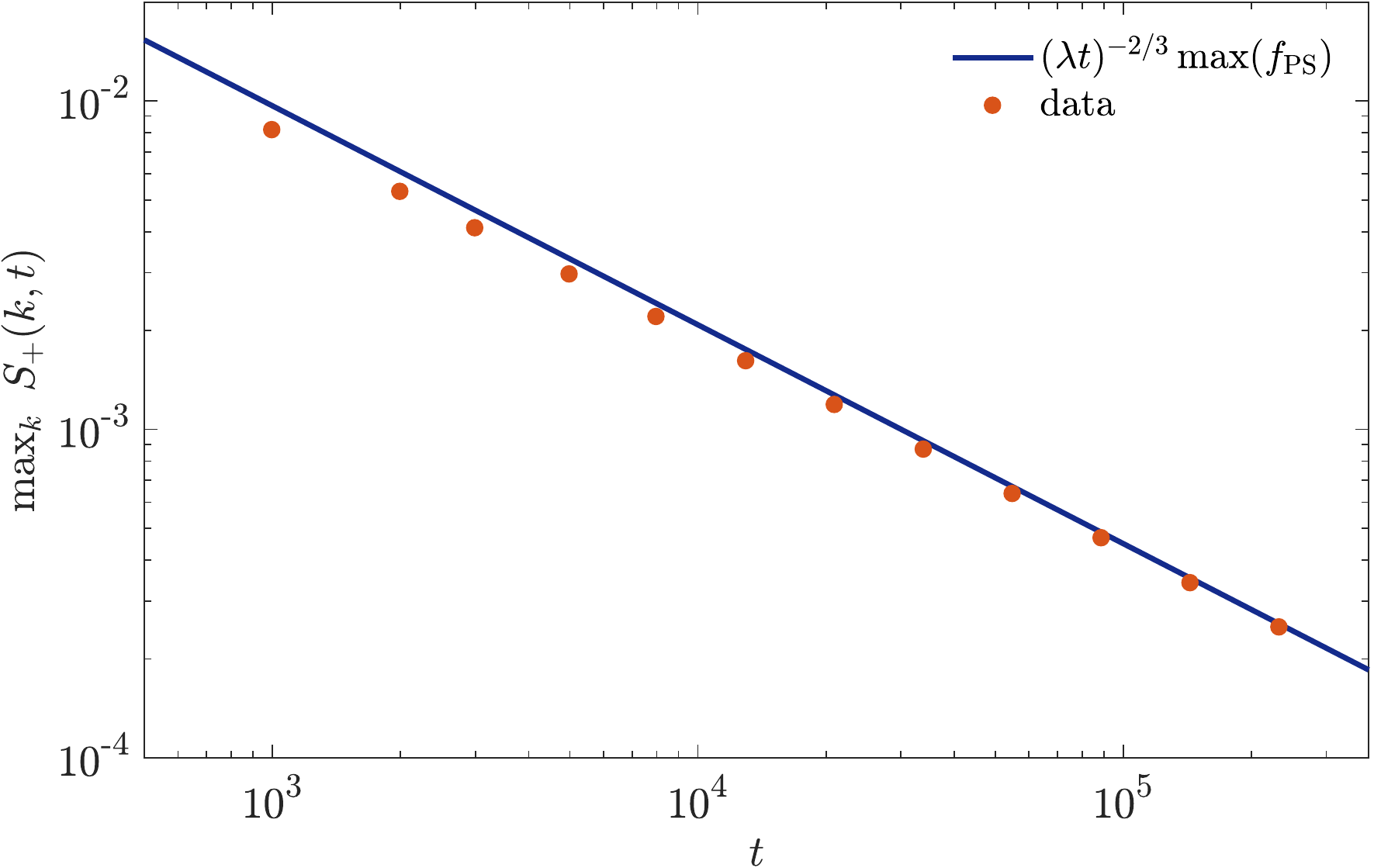}\caption{\label{fig:EX2_log-log-plot-forKPZ-MODE}
Log-log plot for the KPZ-mode
maximum versus time compared to the asymptotic solution Eq~(\ref{eq:KPZ-Mode-Spohn-Asym-Solution}). Statistical errors are in order of symbol size.}
\end{figure}

By looking at a scaling plot  for the full dynamical structure
in position space one notices that the KPZ sound mode does not 
exhibit a good data collapse (\Fref{fig:EX2_Data-collapse-for-KPZ-mode}). 
Also the amplitude at the
maximum shows deviations from the exact theoretical result that
are significantly larger than in the first example 
(\Fref{fig:EX2_log-log-plot-forKPZ-MODE}).

\begin{figure}[H]
\begin{centering}
\includegraphics[height=4.75cm]{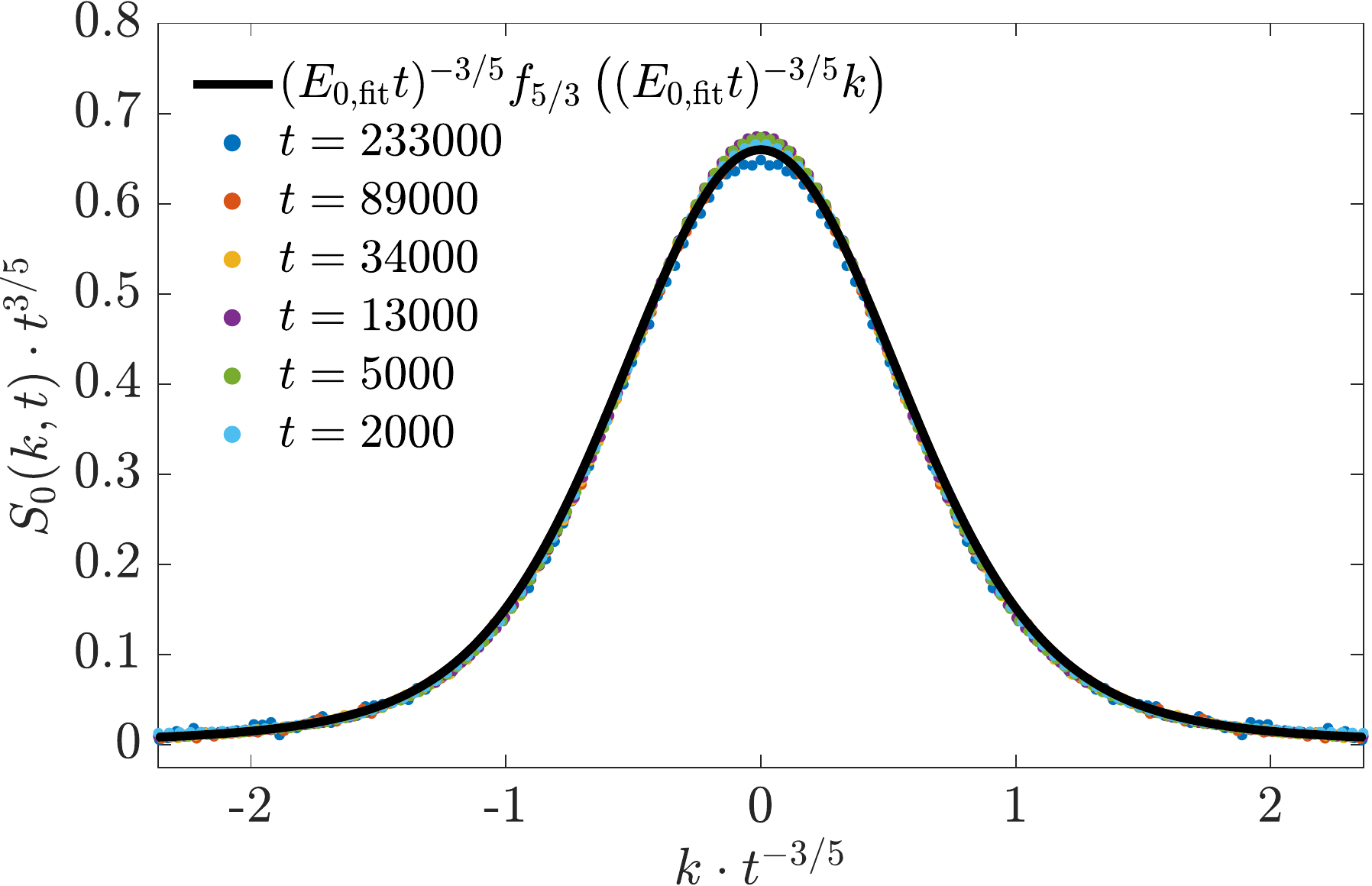}
\includegraphics[height=4.75cm]{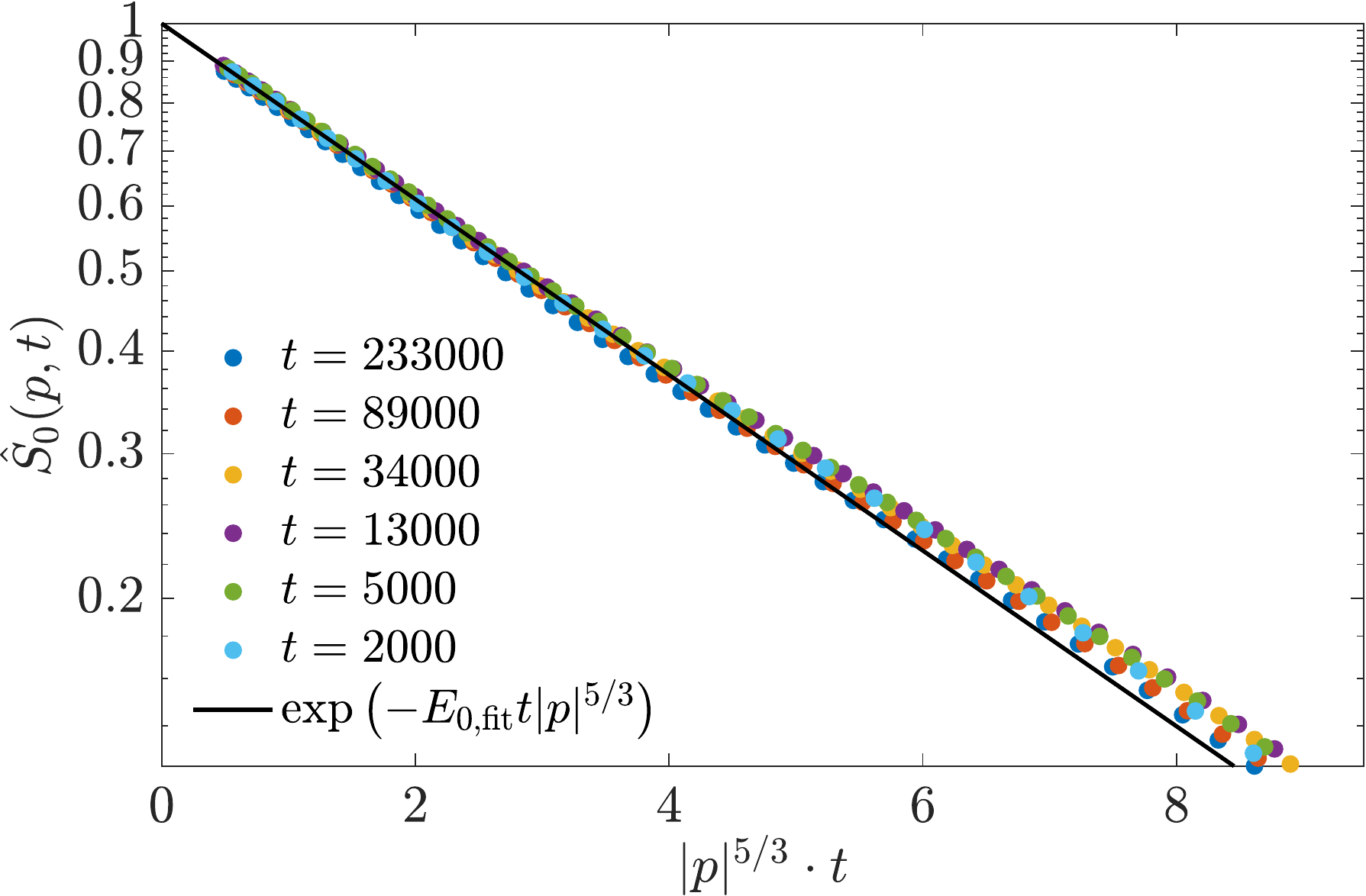}
\par\end{centering}
\caption{\label{fig:EX2_Data-collapse-for-HEAT-mode}
Data collapse in position
space of the $5/3$-L\'evy-mode and scaling-exponent $z=5/3$ compared
to a fitted symmetric $5/3$-L\'evy-stable distribution described
in Fourier representation by eq. (\ref{eq:Heat-Scaling_Fct_DIFF_FINITE_TIME})
and (\ref{eq:heat_scaling_constant}) with $E_{0,\mathrm{fit}}=\left(0.84\mp0.01\right)\cdot E_{0}$
and $E_{0}=0.2927$. Left panel: Position
space, Right panel: Momentum space. 
Statistical errors are in order of symbol
size.}
\end{figure}

\begin{figure}[H]
\begin{centering}
\includegraphics[width=12cm]{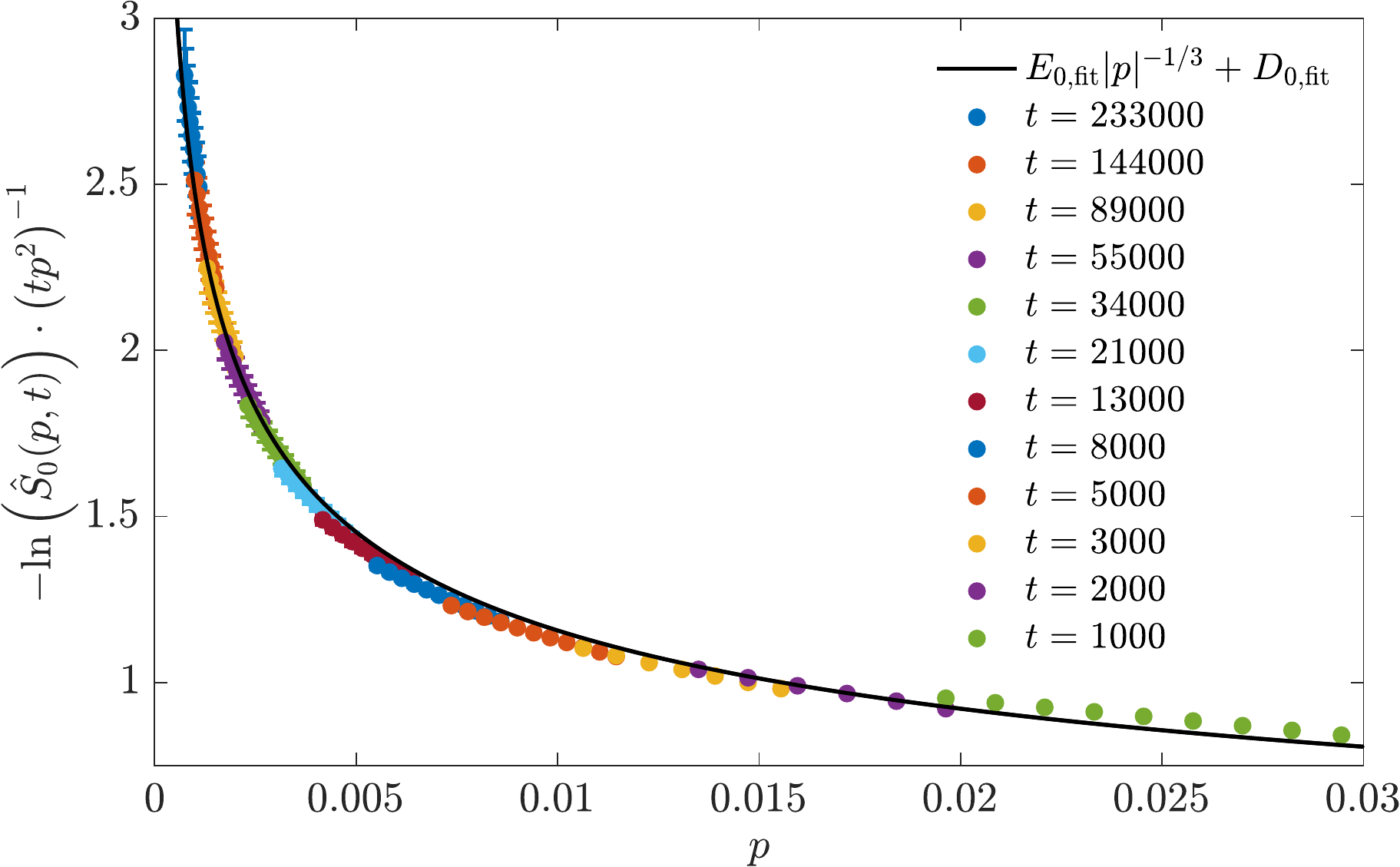}
\par\end{centering}
\caption{\label{fig:EX2_Heat_Mode_diffusive_finite_time_fit}
Fit for diffusive
finite time effects in the $5/3$-L\'evy-mode in fourier-space. Data
is compared to a fitted symmetric $5/3$-stable distribution including
diffusive finite time effects given by eq. (\ref{eq:Heat-Scaling_Fct_DIFF_FINITE_TIME})
with $D_{0,\mathrm{fit}}=0.015\pm0.005$, $E_{0,\mathrm{fit}}=\left(0.84\mp0.01\right)\cdot E_{0}$
and $E_{0}=0.2927$.}
\end{figure}

On the other hand, the L\'evy heat mode exhibits quite good data collapse
(\Fref{fig:EX2_Data-collapse-for-HEAT-mode}). According to the conclusions
drawn from the first example this should be indicative of small diffusive 
corrections. This is confirmed by studying the scaling function
in Fourier space with diffusive corrections. One finds as fit parameters
\begin{eqnarray}
E_{0,\mathrm{fit}} & = & \left(0.84\mp0.01\right)E_{0}\\
D_{0,\mathrm{fit}} & = & 0.015\pm0.005
\end{eqnarray}
corresponding to a small diffusion coefficient. The fit with the
diffusive correction further improves the data collapse, see
\Fref{fig:EX2_Heat_Mode_diffusive_finite_time_fit}.

\section{Conclusions}
The main conclusions from this numerical study can be summarized as
follows.

\begin{itemize}
\item We have found a three-lane lattice gas model that exhibits 
the same stationary fluctuations (two KPZ sound modes with velocities $\pm v
\neq 0$ and L\'evy heat mode with $v=0$) of the three conserved densities 
as generic Hamiltonian dynamics with short-range interaction and conservation 
of mass, energy, and momentum (and no additional conservation laws).
\item The mode coupling matrices of this lattice gas model have the same structure of nonzero elements as mode coupling matrices of generic continuum Hamiltonian dynamics.
\item No fine-tuning of model parameters is required to observe this behaviour.
\item Choosing exclusion dynamics for the lattice gas model allows for 
long-time simulations and very large system size, using a 
canonical ensemble correction scheme introduced here (Sec.~\ref{Subsec:FScorrections})
and which is important to the numerical accuracy of the simulation results.
\item Diffusive corrections generically persist for long times 
but their impact can be varied by changing system parameters.
\end{itemize}

The present work also reconfirms that mode coupling theory 
predicts correctly the predicted universal scaling form
of heat mode, but the precision of non-universal 
scale factor that generally appears in L\'evy modes is still an open problem. 
Also a rigorous mathematical proof of the KPZ/KPZ/5/3-L\'evy scaling for this lattice
model or suitably chosen variants remains a challenge. 
The remarkable rigorous
results of \cite{Bern16} on the 3/2-L\'evy scaling in the case of an anharmonic chain
with two 
conservation laws \cite{Spoh15,Popk15a} provide hope that these questions 
can be answered.

\section*{Acknowledgments}

This research was initiated and supported in part by the International 
Centre for Theoretical Sciences (ICTS) during a visit for participating in the 
program -Non-equilibrium statistical physics (Code: ICTS/Prog-NESP/ 2015/10).
We thank H. Spohn and A. Schadschneider for useful discussions.

\appendix

\section{Computation of the diagonal mode coupling coefficients}

The various constants appearing below are defined in \eref{def:delta} 
($\delta$), \eref{def:kappa} ($\kappa$), \eref{Hessentries} ($x,y,z$),
\eref{vsym} ($v$), and  \eref{def:xi} ($\xi$). We prove \eref{---} by explicit 
computation, beginning with the negative mode
\bea
G^-_{\alpha\alpha} & = & x \left[ R_{-+} \left((R^{-1})_{+\alpha}\right)^2 -
R_{--} \left((R^{-1})_{-\alpha}\right)^2 \right] \nonumber \\
& & + (R^{-1})_{0\alpha} \left[ (z R_{-0} + yR_{-+}) (R^{-1})_{+\alpha} 
- (z R_{-0} + y R_{--}) (R^{-1})_{-\alpha} \right] .
\eea
In particular,
\bea
G^-_{--} & = & x \left[ R_{-+} \left((R^{-1})_{+-}\right)^2 -
R_{--} \left((R^{-1})_{--}\right)^2 \right] \nonumber \\
& & + (R^{-1})_{0-} \left[ (z R_{-0} + yR_{-+}) (R^{-1})_{+-} 
- (z R_{-0} + y R_{--}) (R^{-1})_{--} \right] \\
& = & \frac{x}{\xi^3} \left[ - \frac{b\sqrt{\kappa_0}}{v+\delta}\left(\frac{b^2\sqrt{\kappa_0}\kappa}{v+\delta}\right)^2 -
\frac{b\sqrt{\kappa_0}}{v-\delta} \left(\frac{b^2\sqrt{\kappa_0}\kappa}{v-\delta}\right)^2 \right] \nonumber \\
& & + \frac{\sqrt{\kappa_0}}{\xi^3} \left[ - \left(\frac{z}{\sqrt{\kappa_0}} - \frac{by\sqrt{\kappa_0}}{v+\delta}\right) \frac{b\sqrt{\kappa_0}\kappa}{v+\delta} 
- \left(\frac{z}{\sqrt{\kappa_0}} + \frac{by\sqrt{\kappa_0}}{v-\delta}\right) \frac{b\sqrt{\kappa_0}\kappa}{v-\delta} \right] \\
& = & - \frac{v}{b\xi^3\sqrt{\kappa_0}} \left[ x
\left( \frac{1}{2} + \frac{\delta^2}{b^2\kappa\kappa_0}\right) 
+ \frac{y\delta }{b\kappa}
+ z \right],
\eea

\bea
G^-_{00} & = & x \left[ R_{-+} \left((R^{-1})_{+0}\right)^2 -
R_{--} \left((R^{-1})_{-0}\right)^2 \right] \nonumber \\
& & + (R^{-1})_{00} \left[ (z R_{-0} + yR_{-+}) (R^{-1})_{+0} 
- (z R_{-0} + y R_{--}) (R^{-1})_{-0} \right] \\
& = & \frac{x}{\xi^3} \left[ - \frac{b\sqrt{\kappa_0}}{v+\delta} \kappa -
\frac{b\sqrt{\kappa_0}}{v-\delta} \kappa \right] \nonumber \\
& & - \frac{\delta}{b\sqrt{\kappa}\xi^3} \left[ \left(\frac{z}{\kappa} - \frac{by \sqrt{\kappa_0}}{v+\delta}\right) \sqrt{\kappa}
- \left(\frac{z}{\kappa}   + \frac{by\sqrt{\kappa_0}}{v-\delta}\right) \sqrt{\kappa} \right] \\
& = & -\frac{v}{b\xi^3\sqrt{\kappa_0}} \left( x - \frac{y\delta}{b\kappa}\right),
\eea
and
\bea
G^-_{++} & = & x \left[ R_{-+} \left((R^{-1})_{++}\right)^2 -
R_{--} \left((R^{-1})_{-+}\right)^2 \right] \nonumber \\
& & + (R^{-1})_{0+} \left[ (z R_{-0} + yR_{-+}) (R^{-1})_{++} 
- (z R_{-0} + y R_{--}) (R^{-1})_{-+} \right] \\
& = & \frac{x}{\xi^3} \left[ - \frac{b\sqrt{\kappa_0}}{v+\delta} \left(\frac{b\sqrt{\kappa_0}\kappa}{v-\delta}\right)^2 -
\frac{b\sqrt{\kappa_0}}{v-\delta} \left(\frac{b\sqrt{\kappa_0}\kappa}{v+\delta}\right)^2 \right] \nonumber \\
& & + \frac{\sqrt{\kappa_0}}{\xi^3} \left[ \left(\frac{z}{\sqrt{\kappa_0}} - \frac{by\sqrt{\kappa_0}}{v+\delta}\right) \frac{b\sqrt{\kappa_0}\kappa}{v-\delta} 
+ \left(\frac{z}{\sqrt{\kappa_0}} + \frac{by\sqrt{\kappa_0}}{v-\delta}\right) \frac{b\sqrt{\kappa_0}\kappa}{v+\delta} \right] \\
& = & - \frac{v}{b\xi^3\sqrt{\kappa_0}} \left( \frac{x}{2} - z \right)
\eea

We have for mode 0
\bea
G^0_{\alpha\alpha} 
& = & x \left[ R_{0+} \left((R^{-1})_{+\alpha}\right)^2 -
R_{0-} \left((R^{-1})_{-\alpha}\right)^2 \right] \nonumber \\
& & + (R^{-1})_{0\alpha} \left[ (z R_{00} + yR_{0+}) (R^{-1})_{+\alpha} 
- (z R_{00} + y R_{0-}) (R^{-1})_{-\alpha} \right] .
\eea
This yields
\bea
G^0_{--} 
& = & x \left[ R_{0+} \left((R^{-1})_{+-}\right)^2 -
R_{0-} \left((R^{-1})_{--}\right)^2 \right] \nonumber \\
& & + (R^{-1})_{0-} \left[ (z R_{00} + yR_{0+}) (R^{-1})_{+-} 
- (z R_{00} + y R_{0-}) (R^{-1})_{--} \right] \\
& = & \frac{x}{\xi^3} \left[ \frac{1}{\sqrt{\kappa}} 
\left( \frac{b\sqrt{\kappa_0}\kappa}{v+\delta} \right)^2 -
\frac{1}{\sqrt{\kappa}} \left(\frac{b\sqrt{\kappa_0}\kappa}{v-\delta}\right)^2 \right] \nonumber \\
& & + \frac{ \sqrt{\kappa_0}}{\xi^3} \left[ \left( \frac{z\delta}{b\sqrt{\kappa}\kappa_0} - 
\frac{y}{\sqrt{\kappa}}\right) \frac{b\sqrt{\kappa_0}\kappa}{v+\delta} 
- \left( -\frac{z\delta}{b\sqrt{\kappa}\kappa_0} + 
\frac{y}{\sqrt{\kappa}}\right) \frac{b\sqrt{\kappa_0}\kappa}{v-\delta} \right] \\
& = & - \frac{v}{b\xi^3\sqrt{\kappa}}
\left((x-z)\frac{\delta}{b\kappa_0} + y \right).
\eea
The expressions given in \eref{---} then follow from \eref{def:xi}.

\end{document}